\documentclass[a4paper,twoside,openany,11pt]{memoir} 
\usepackage{minibox}
\usepackage{longtable}
\def\fullheadfoot{0}

\usepackage[english]{babel}
\usepackage{microtype,xspace}
\usepackage[utf8]{inputenc}	

\usepackage{amsfonts,amsmath,amssymb,bm,mathrsfs,mathtools,dsfont} 	
\allowdisplaybreaks 	
\usepackage[table,dvipsnames]{xcolor}
\usepackage{hyperref}
\usepackage{slashed}
\usepackage{multicol}

\usepackage{siunitx}
\usepackage{geometry}
\usepackage[sort&compress,numbers,merge]{natbib}
\usepackage{chapterbib}
\addto\captionsenglish{\renewcommand*{\bibname}{References}}
\makeatletter
\renewcommand{\@memb@bchap}{ 
\bibmark \prebibhook
}
\makeatother

\usepackage{graphicx}
\graphicspath{{./Figures/}{./Tikzfig/}}
\usepackage{caption,subcaption}
\usepackage{multirow}
\renewcommand{\arraystretch}{1.2}
\usepackage{tabularx}
\newcolumntype{Y}{>{\centering\arraybackslash}X}

\usepackage[shortlabels]{enumitem}
\setlist{itemsep=.1em,topsep=.5em}
\SetEnumerateShortLabel{i}{\textit{\roman*}}

\definecolor{red}{rgb}{0.6,.0706,.1373}
\definecolor{blue}{rgb}{0,0.396,0.741}
\definecolor{green}{rgb}{0.25,0.6,0.2}
\definecolor{teal}{rgb}{0.11,0.6,0.6}
\definecolor{orange}{rgb}{.8, .4806, 0.173}
\definecolor{yellow}{rgb}{.8, .7, 0.05}
\colorlet{blueref}{blue!80!black}
\colorlet{bluelink}{blue!90!black}
\hypersetup{
	colorlinks, 
	bookmarksopen, 
	bookmarksnumbered,
	citecolor=blueref, 		
	linkcolor=bluelink,	
	urlcolor=bluelink,			
	pdftitle={A Proof of Concept for Matchete},    
}

\addto\captionsenglish{}

\renewcommand{\contentsname}{Contents}
\renewcommand{\printtoctitle}[1]{}

\settocdepth{subsection}
\newcommand{\toc}{ {
	\hypersetup{linkcolor = black} 
	\vspace*{-.06\textheight}	
	\tableofcontents*
	\thispagestyle{standardstyle} 
} }

\makeatletter
\newcommand*\ifthispageodd{%
  \checkoddpage
  \ifoddpage
    \expandafter\@firstoftwo
  \else
    \expandafter\@secondoftwo
  \fi
}
\makeatother

\usepackage[noindentafter]{titlesec} 
\counterwithout{section}{chapter} 
\counterwithout{figure}{chapter} 
\counterwithout{table}{chapter} 
\numberwithin{equation}{section} 
\setsecnumdepth{subsubsection}

\DeclareMathVersion{sans}
\SetSymbolFont{operators}{sans}{OT1}{cmbr}{m}{n}
\SetSymbolFont{letters}{sans}{OML}{cmbrm}{m}{it}
\SetSymbolFont{symbols}{sans}{OMS}{cmbrs}{m}{n}
\SetMathAlphabet{\mathit}{sans}{OT1}{cmbr}{m}{sl}
\SetMathAlphabet{\mathbf}{sans}{OT1}{cmbr}{bx}{n}
\SetMathAlphabet{\mathtt}{sans}{OT1}{cmtl}{m}{n}
\SetSymbolFont{largesymbols}{sans}{OMX}{iwona}{m}{n}

\DeclareMathVersion{boldsans}
\SetSymbolFont{operators}{boldsans}{OT1}{cmbr}{b}{n}
\SetSymbolFont{letters}{boldsans}{OML}{cmbrm}{b}{it}
\SetSymbolFont{symbols}{boldsans}{OMS}{cmbrs}{b}{n}
\SetMathAlphabet{\mathit}{boldsans}{OT1}{cmbr}{b}{sl}
\SetMathAlphabet{\mathbf}{boldsans}{OT1}{cmbr}{bx}{n}
\SetMathAlphabet{\mathtt}{boldsans}{OT1}{cmtl}{b}{n}
\SetSymbolFont{largesymbols}{boldsans}{OMX}{iwona}{bx}{n}

\titleformat{\section}{\centering \Large \bfseries \sffamily \mathversion{boldsans} \color{blue!80!black} }{\thesection}{15pt}{}{}
\titlespacing{\section}{0pt}{15pt}{5pt}
\titleformat{\subsection}{\large \sffamily \mathversion{sans} \color{blue!70!black} }{\thesubsection}{10pt}{}{}
\titlespacing{\subsection}{0pt}{10pt}{5pt}
\titleformat{\subsubsection}{\normalsize \sffamily \itshape \mathversion{sans} \color{blue!70!black} }{\thesubsubsection}{10pt}{}{}
\titlespacing{\subsubsection}{0pt}{10pt}{0pt}

\ifnum\fullheadfoot=1
	\makepagestyle{standardstyle}
	\makeoddhead{standardstyle}{\sffamily \mathversion{subsectionmath} \rightmark}{}{\sffamily \bfseries{\thepage}}
	\makeevenhead{standardstyle}{\sffamily \bfseries{\thepage}}{}{\sffamily \mathversion{subsectionmath} \leftmark}
	\makeheadrule{standardstyle}{\textwidth}{\normalrulethickness}
	\makepsmarks{standardstyle}{
		\nouppercaseheads
		\createmark{section}{both}{shownumber}{\ }{\hspace{1.2em}  }
		\createmark{subsection}{right}{shownumber}{}{\hspace{1.2em} }
	}
	
	\copypagestyle{appstyle}{standardstyle}
	\makeevenhead{appstyle}{\sffamily \bfseries{\thepage}}{}{ \sffamily \mathversion{subsectionmath} \leftmark}
	\makepsmarks{appstyle}{
		\nouppercaseheads
		\createmark{section}{both}{nonumber}{\ }{\ \ }
		\createmark{subsection}{right}{shownumber}{}{\ \ }
	}
\else 
	\makepagestyle{standardstyle}
	\makeoddfoot{standardstyle}{}{\sffamily -- {\bfseries\thepage} --}{} 
	\makeevenfoot{standardstyle}{}{\sffamily -- {\bfseries\thepage} --}{}
	\copypagestyle{appstyle}{standardstyle}
\fi

\createplainmark{toc}{both}{\contentsname}
\createplainmark{bib}{both}{\bibname}

\makeatletter
\let\MyIntOrig\int
\def\MyIntSpace{\hspace{-.35em}} 
\def\int{\MyInt}
\def\MyInt{\MyIntOrig\MyIntSkipMaybe}
\def\MyIntSkipMaybe{
	\@ifnextchar_{\MyIntSkipScript}{%
		\@ifnextchar^{\MyIntSkipScript}{%
			\@ifnextchar\limits{\MyIntSkipTok}{%
				\@ifnextchar\nolimits{\MyIntSkipTok}{%
					\MyIntSpace}}}}%
}
\def\MyIntSkipScript#1#2{#1{#2}\MyIntSkipMaybe}
\def\MyIntSkipTok#1{#1\MyIntSkipMaybe}

\newcommand{\pushright}[1]{\ifmeasuring@#1\else\omit\hfill$\displaystyle#1$\fi\ignorespaces}
\makeatother

\usepackage{bbm}
\RequirePackage{fix-cm}
\usepackage{fontawesome}
\usepackage{booktabs}
\usepackage{multirow}
\usepackage{soul}
\usepackage{bbold}

\newcommand{\tr}{\mathop{\mathrm{tr}} }
\newcommand{\eminus}{\vcenter{\hbox{\scalebox{0.6}[1]{$ - $}}}}	

\newcommand{\hc}{\; + \; \mathrm{h.c.} \;}

\newcommand{\diag}{\mathop{\mathrm{diag}}}
\newcommand{\rep}[1]{\mathbf{#1}}

\newcommand{\sscript}[1]{{\scriptscriptstyle \mathrm{#1}}}
\newcommand{\eom}[1]{\mathcal{D}\left( #1 \right)}

\renewcommand{\L}{\mathcal{L}}
\newcommand{\LL}{\mathrm{L}}
\newcommand{\RR}{\mathrm{R}}
\newcommand{\U}{\mathrm{U}}
\newcommand{\SU}{\mathrm{SU}}

\newcommand{\UV}{\sscript{UV}}
\newcommand{\EFT}{\sscript{EFT}}

\newcommand{\msbar}{$ \overline{\text{\small MS}} $\xspace}

\newcommand{\matchete}{\texttt{Matchete}\xspace}

\definecolor{verde}{cmyk}{0.92,0,0.59,0.25}
\newcommand{\pkg}[1]{{\texttt{#1}}\xspace}

\usepackage[utf8]{inputenc}

\definecolor{DarkGray}{gray}{0.40}
\usepackage{mmacells}
\mmaSet{
  morefv={gobble=2},
  linklocaluri=mma/symbol/definition:#1,
  morecellgraphics={yoffset=1.9ex},
  moredefined={DefineGaugeGroup, DefineField, DefineCoupling,
    Fermion, Scalar, Charges, Mass, SelfConjugate, Heavy, 
    H, U1, SU, fund, eps, CG, l, EFTOrder, Indices, LoopOrder, ModelParameters,
    FreeLag, NiceForm, PlusHc, Bar, PR, PL, Match, GreensSimplify, EOMSimplify, LoadModel,
    Match, CovariantLoop, HcSimplify, Contract, Light, m, SelectOperatorClass,
    FundAlphabet, AdjAlphabet, IndexAlphabet, DefineFlavorIndex, Chiral, LeftHanded, RightHanded, RelabelIndices, CheckLagrangian
  }
}

\begin{document}

\thispagestyle{empty}
\renewcommand*{\thefootnote}{\fnsymbol{footnote}}
\begin{center} 
\begin{minipage}{15.5cm}
\vspace{-0.7cm}
\begin{flushright}
{\footnotesize \itshape
MITP-22-105\\
TUM-HEP-1443/22\\[-3pt]
ZU-TH-58/22
}
\end{flushright}


\end{minipage}
\end{center}

\begin{center}
	{\sffamily \bfseries \fontsize{16.5}{20}\selectfont \mathversion{boldsans}
	A Proof of Concept for Matchete:\\ An Automated Tool for Matching Effective Theories\\[-.5em]
	\textcolor{blue!80!black}{\rule{0.95\textwidth}{2pt}}\\[1ex]
	\quad\includegraphics[width=.8\textwidth]{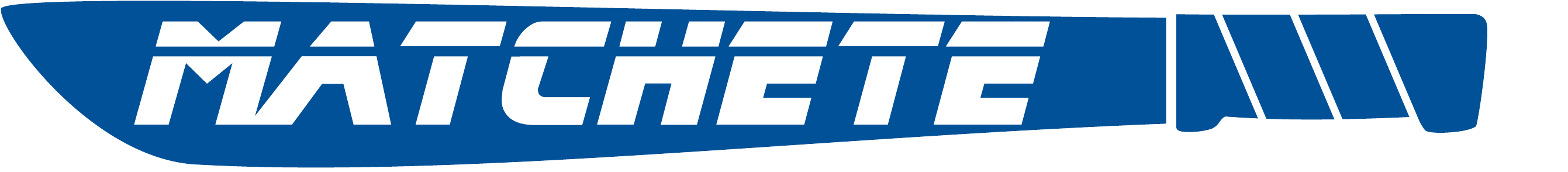}\\
	\vspace{.03\textheight}}
	{\sffamily \mathversion{sans} \Large Javier Fuentes-Mart\'{\i}n,$^{1,2}$\footnote{javier.fuentes@ugr.es} 
	Matthias König,$^{3}$\footnote{matthias.koenig@tum.de}
	Julie Pagès,$^{4}$\footnote{julie.pages@physics.ucsd.edu}\\[5pt]
	Anders Eller Thomsen,$^{5}$\footnote{thomsen@itp.unibe.ch}
	and Felix Wilsch$^{6}$\footnote{felix.wilsch@physik.uzh.ch}}
	\\[1.25em]
	{ \small \sffamily \mathversion{sans} 
	$^{1}\,$  Departamento de F\'isica Te\'orica y del Cosmos, Universidad de Granada,\\
    Campus de Fuentenueva, E–18071 Granada, Spain\\[5pt]
    $^{2}\,$  PRISMA Cluster of Excellence \& Mainz Institute for Theoretical Physics,\\
    Johannes Gutenberg University, D-55099 Mainz, Germany\\[5pt]
	$^{3}\,$ Physik Department T31, Technische Universit\"at M\"unchen,\\
	James-Franck-Str. 1, D-85748 Garching, Germany\\[5pt]
	$^{4}\,$ Department of Physics, University of California at San Diego, \\9500 Gilman Drive,
    La Jolla, CA 92093-0319, USA\\[5pt]
	$^{5}\,$ Albert Einstein Center for Fundamental Physics, Institute for Theoretical Physics,\\ University of Bern, CH-3012 Bern, Switzerland\\[5pt]
	$^{6}\,$ Physik-Institut, Universit\"at Z\"urich, CH-8057 Z\"urich, Switzerland
	}
	\\[.005\textheight]{\itshape \sffamily \today}
	\\[.0\textheight]
\end{center}
\setcounter{footnote}{0}
\renewcommand*{\thefootnote}{\arabic{footnote}}%
\suppressfloats	

\begin{abstract} 
Studying the impact of new-physics models on low-energy observables necessitates matching to effective field theories at the relevant mass thresholds. We introduce the first public version of \matchete, a computer tool for matching weakly-coupled models at one-loop order. It uses functional methods to directly compute all matching contributions in a manifestly gauge-covariant manner, while simplification methods eliminate redundant operators from the output. We sketch the workings of the program and provide examples of how to match simple Standard Model extensions. The package, documentation, and example notebooks are publicly available at
\url{https://gitlab.com/matchete/matchete}.
\end{abstract}

\newpage
\noindent\textcolor{blue!80!black}{\rule{1\textwidth}{1.5pt}}
\vspace*{-.035\textheight}
\subsection*{Table of Contents}
\vspace*{-.02\textheight}
\toc
\noindent\textcolor{blue!80!black}{\rule{1\textwidth}{1.5pt}}

\section{Introduction}
\label{sec:intro}
The advent of the LHC heralded a new era for beyond-the-Standard-Model (BSM) physics. With the discovery of the Higgs boson and no direct signs of new resonances, we see indications of a mass gap up to the scale of yet-to-be-discovered new physics (NP). The focus of the community is shifting to precision flavor and electroweak physics in order to search for indirect signs of new particles and potentially probe scales far beyond the reach of resonance searches. The result has been a renaissance of Effective Field Theories (EFTs) applied to BSM physics often using the Standard Model Effective Theory (SMEFT), whose basis was first determined after the LHC went into service~\cite{Grzadkowski:2010es}. The use of EFTs goes all the way back to Fermi's theory and has long since reached maturity within the Standard Model (SM), facilitating SM predictions for many precision observables. Now, new methods are rapidly being developed for BSM physics with an aspiration of reaching a similar level of maturity. The new challenge to achieving this goal is the need for a near-complete level of generality, as the nature of NP has yet to be revealed.

To determine the low-energy effects of high-scale NP, one typically has to perform sequential matching to consecutive EFTs at the relevant mass thresholds and renormalization group~(RG) running between these scales. In the absence of any light new particles, the running and matching machinery is already available to handle computations below the NP mass threshold: the one-loop RG equations in the SMEFT~\cite{Jenkins:2013zja, Jenkins:2013wua, Alonso:2013hga, Alonso:2014zka}, the matching to the Low-Energy Effective Theory (LEFT) at the weak scale~\cite{Aebischer:2015fzz,Jenkins:2017jig, Dekens:2019ept}, and the LEFT RG equations~\cite{Jenkins:2017dyc} have been determined and even implemented in computational tools~\cite{Celis:2017hod, Aebischer:2018bkb, Fuentes-Martin:2020zaz}. Many tools are also available for phenomenological analyses of theories within the SMEFT and LEFT frameworks~\cite{Brivio:2017btx,GAMBITFlavourWorkgroup:2017dbx,Aebischer:2018iyb,Straub:2018kue,Brivio:2019irc,DeBlas:2019ehy,Dedes:2019uzs,Hartland:2019bjb,Uhlrich:2020ltd,Ellis:2020unq,EOSAuthors:2021xpv,Allwicher:2022mcg}. 

The sticking point for a long time has been performing the matching computation of BSM models to their EFTs. Although it is tempting to think of the target EFT as the SMEFT, we should bear in mind that realistic BSM constructions can contain a rich NP sector spanning large ranges of energy scales, calling for intermediate-scale EFTs. Alternatively, the presence of additional light fields, for example axion-like or dark-matter particles, demands extensions of the SMEFT~(see, e.g.,~\cite{Criado:2021trs,Aebischer:2022wnl,Chala:2020wvs,Bauer:2020jbp,Galda:2021hbr}). The unclear nature of both the UV model and the target EFT makes matching a formidable task. Functional methods promise a direct approach to the problem~\cite{Gaillard:1985uh,Chan:1986jq,Cheyette:1987qz,Chan:1985ny,Fraser:1984zb,Aitchison:1984ys,Aitchison:1985pp,Aitchison:1985hu,Cheyette:1985ue,Chan:1986jq,Dittmaier:1995cr,Dittmaier:1995ee,Henning:2014wua,Drozd:2015rsp,Henning:2016lyp,Fuentes-Martin:2016uol,Zhang:2016pja,Cohen:2020fcu,delAguila:2016zcb,Boggia:2016asg,Dittmaier:2021fls}. They entirely circumvent the matching of individual amplitudes and produce the EFT Lagrangian directly, albeit unsimplified, without requiring any prior knowledge about its structure or symmetries. The method has produced general results in the form of the Universal One-Loop Effective Action~\cite{Ellis:2016enq, Ellis:2017jns, Kramer:2019fwz, Angelescu:2020yzf, Ellis:2020ivx}, several tools to assist part of the matching computations~\cite{Criado:2017khh,Cohen:2020qvb,Fuentes-Martin:2020udw,Bakshi:2018ics}, and has been used for a number of simple BSM models~\cite{Zhang:2021jdf,Dedes:2021abc,Du:2022vso,Li:2022ipc,Liao:2022cwh,Guedes:2022cfy}. Nevertheless, the package we present here represents the first truly automated, end-to-end one-loop matching tool based on functional methods, making it competitive with the diagrammatic \texttt{matchmakereft}~\cite{Carmona:2021xtq} but with the advantages of the functional approach. Thanks to these new tools, fast and competent matching requiring little more than the press of a button is finally becoming feasible. Not only that, matching tools can easily be repurposed to compute RG equations for other EFTs, as both types of computations require the evaluation of loop integrals in the hard region. 

Here we introduce a first public, proof-of-concept version (\texttt{v0.1.0}) of the \pkg{Mathematica} package \matchete---Matching Effective Theories Efficiently---to solve the problem of matching weakly-coupled UV models to their EFTs at the relevant mass thresholds. It uses functional methods~\cite{Fuentes-Martin:2016uol,Zhang:2016pja,Cohen:2020fcu}, which facilitates direct matching without the need for specifying a target basis for the EFT. This feature is especially useful in theories that match into EFTs other than the SMEFT or when extending EFT matching beyond dimension-six operators. The automated application of these methods was previously demonstrated by the authors in the \texttt{SuperTracer} package~\cite{Fuentes-Martin:2020udw}, which \matchete supersedes. Furthermore, we make significant headway with the challenging task of automatically simplifying the EFT Lagrangian to an on-shell basis. The design of the package includes a simple and user-friendly interface that considerably simplifies the user input while still allowing for very generic implementations. In essence, the user can write down the Lagrangian in a \pkg{Mathematica} notebook, in manner that is very close to a pen-and-paper form, and leave the rest to the package. While there are still many features and capabilities that we would like to implement over the next years, this proof-of-concept release already represents a major leap in the development of (functional) matching tools and can greatly assist many matching computations, including those in multiple BSM scenarios. The limitations in this release are reflected in the discussion of the future prospects of the package in Section~\ref{sec:prospects}. 

This paper is meant as a short introduction to give a flavor of the first public version of \matchete. The paper contains a brief description of the underlying package structure and gives some hands-on examples of how to use it. It is not a comprehensive guide to the use of the program. For more detailed instructions, the user is encouraged to consult the documentation notebook included with the distribution. Section~\ref{sec:organization} presents the organization and use of the package in broad scopes, touching on the specific methods used in the computation. This is followed with some concrete usage example in Section~\ref{sec:examples}, including simple extensions of the SM, to give the reader a feel for the practical applications. We conclude the paper in Section~\ref{sec:prospects} with a short discussion of the direction of future package developments.

\section{Organization of the Package} \label{sec:organization}
\begin{figure}
        \centering
        \includegraphics{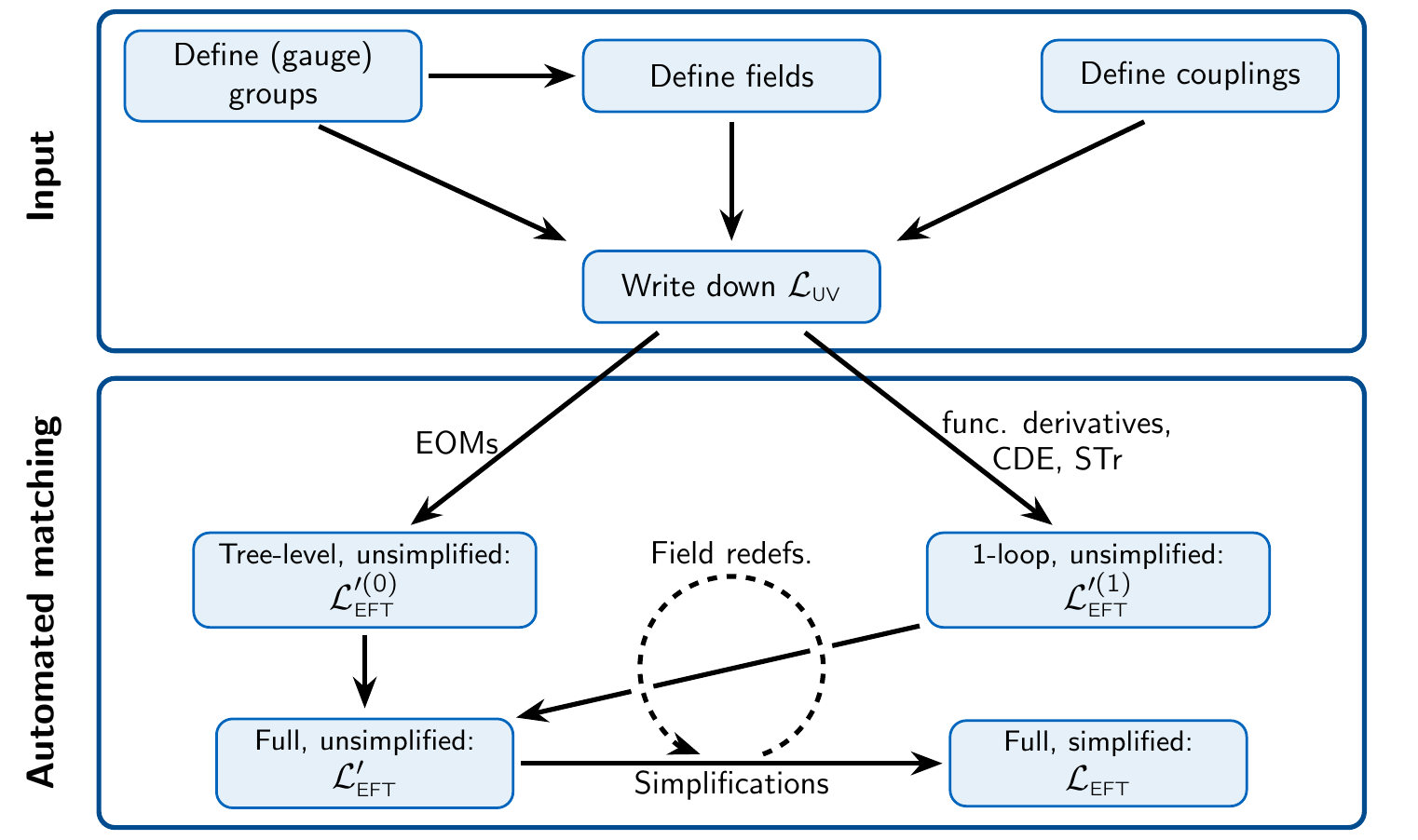}
        \caption{Schematic representation of the \matchete workflow. The user has to specify the (gauge) groups, fields, and couplings of a UV model before writing down the Lagrangian. This can be passed to functions for EFT matching at tree- and one-loop level. Simplification methods with identities and field redefinitions can then reduce the EFT Lagrangian.}
        \label{fig:matchete_diagram} 
    \end{figure}

\matchete is organized around Lagrangians and operators, which are the objects the user will interact with in the workflow illustrated in Figure~\ref{fig:matchete_diagram}. With standard functional methods, a UV Lagrangian is matched to an EFT Lagrangian at tree- and one-loop level. However, by nature of the functional approach, the EFT Lagrangian is not simplified and contains many redundant operators that need to be reduced to a basis. A core part of \matchete consists of powerful methods for simplifying the matching result to a (near-)basis to bridge the gap to a useful EFT Lagrangian. Various functions are also available for manipulating the output in various ways and identifying individual contributions. The simplification methods do not handle evanescent contributions yet, so the output is in a $ d $-dimensional basis, which is redundant in a physical renormalization scheme~\cite{Fuentes-Martin:2022vvu}.

\subsection{Model setup and internal representations }
Much of the user experience invariably concerns the input of models into the program. First, one must specify what (gauge) groups and representations are available for objects to transform under before one can specify what manner of fields and couplings are involved. Only then can a Lagrangian be written down. To achieve this, \matchete contains specific methods for Lorentz contractions and Dirac algebra.

\subsubsection{Symmetry groups}
All manner of group invariants show up in quantum-field-theory computations. These can be as simple as Kronecker deltas or generators of a representation, or they can be much more complicated once more exotic representations are involved. We will refer to all such invariant tensors as Clebsch--Gordan (CG) coefficients in line with the well-known $ \SU(2) $ case. We are unaware of analytic rules for evaluating generic contractions of CG coefficients that apply to all cases, so a more constructive approach is used in \matchete.

\matchete contains a module for handling all things related to group and representation theory, to allow for generic gauge and/or symmetry groups. Upon specifying a simple Lie group, the module can determine the weights, dimensions, and other information regarding the representations with standard methods.\footnote{We found Ref.~\cite{Cahn:1985wk} very useful as a primer and reminder for relevant Lie algebra methods.} To determine CG coefficients, which describe how to combine weights from multiple representations in an invariant manner, we implemented the algorithm of Refs.~\cite{SousadaFonseca:2013hlw,Fonseca:2020vke}, which casts the problem in terms of linear algebra. With this method, the program can explicitly construct the CG tensors.
  
When the user specifies a simple (gauge) group to be included in a model, \matchete automatically generates several common representations and CG coefficients and more can be initialized by the user with build-in routines. The CGs are referred to symbolically for all input and output purposes. However, when contracting CG objects, the symbols are replaced internaly with numerical tensors, contracted, and, finally, projected to a basis of CG coefficients. This procedure allows for efficient evaluation of CG products, with a minimum of inconvenience to the user.

\subsubsection{Fields and couplings}
All objects in a Lagrangian have properties associated with them that are necessary for determining what algebraic manipulations are possible. The field and coupling objects appearing in \matchete Lagrangians carry most of this information with them as they are passed along to various routines. Although concise, the amount of information contained in a Lagrangian, or even an operator, is considerable, which can be useful for careful manipulations of the output should the user so desire. In most common cases, the user will want to exploit the convenience of the notebook format to view the output in a more legible form, and build-in routines allow the user to print \matchete objects in the format of a regular textbook.

Despite all the information contained in field objects, users can simply refer to them with their label (name) and the indices they might have. All that is required for the user is to define the properties of a field: spin, mass, flavor and gauge representations, and whether if it self-conjugate (real for scalars, Majorana for fermions). Gauge fields are even easier to implement, as they are automatically defined with their gauge group. Couplings, similarly, need to be defined beforehand to specify their flavor indices and mass dimensions.

\subsection{Matching step}
The input UV Lagrangian is matched to an EFT under the assumption that all heavy masses are of the same order $ M_a \sim \Lambda $ (otherwise, the matching will have to be performed sequentially accounting for RG running), which sets the heavy scale of the problem. This allows for arranging the EFT as a double expansion in the heavy scale and the loop order:
    \begin{equation}
    \L_\EFT = \sum_{\ell=0} \sum_{n=4} \dfrac{\hbar^\ell}{(4\pi)^{2\ell} \Lambda^{n-4}} \L_\EFT^{(\ell, n)}\,.
    \end{equation}
\matchete features routines for computing $ \L_\EFT^{(0, n)} $ and $ \L_\EFT^{(1, n)} $. There is no fundamental obstacle preventing the evaluation of higher-dimensional terms with the current implementation, although limits of computing power make pushing beyond dimension-six for one-loop terms time intensive. In practice, the mass expansion is performed in terms of the light dimensions, counting the canonical dimension of light fields, covariant derivatives, and light masses/dimensional couplings.

\subsubsection{Tree level}
Matching at tree level comes down to solving the equations of motion (EOMs) of the heavy fields as it is commonly done by hand. This approach has also been applied to automated tree-level matching in \texttt{MatchingTools}~\cite{Criado:2017khh}. Schematically, with heavy fields~$ \Phi_a $ and light fields~$ \phi $, the UV action is $ S_\UV[\Phi, \phi]= \int_x \, \tfrac{1}{2} \Phi_a \Delta_{ab}^{\eminus 1}(D,\, M) \Phi_b + S_\mathrm{int} $, where $ \Delta^{\eminus 1}$ is the appropriate kinetic operator for the heavy fields and $S_\mathrm{int}$ is the interacting part of the action (including both heavy and light fields). The solution to the heavy-field EOMs in the presence of light fields reads
    \begin{equation}
    \hat{\Phi}_a[\phi] = - \Delta_{ab} \dfrac{\delta S_\mathrm{int}}{\delta \Phi_b} \big[\hat{\Phi}[\phi],\, \phi\big].
    \end{equation}
\matchete is equipped with routines to take functional derivatives of the action after which this equation can be solved iteratively order by order in the mass expansion. With the solution in hand, the tree-level EFT is given by 
    \begin{equation}
    S_\EFT^{(\ell = 0)}[\phi] = S_\UV \big[\hat{\Phi}[\phi],\, \phi\big].
    \end{equation}
An efficient truncation of higher-order terms ensures excellent performance of this method.

\subsubsection{One-loop level}
One-loop contributions to the EFT encode the high-energy components of one-loop effects in the UV theory. In the functional formalism, there is but a single functional topology at one-loop order, which is captured by a \textit{supertrace}---a generalization of the functional trace that accounts for the presence of mixed bosonic and fermionic objects. The key object to consider is the fluctuation operator 
    \begin{equation}\label{eq:flucOp}
    \frac{\delta^2 S_\UV}{\delta\eta_j\delta\bar\eta_i}\big[\hat{\Phi}[\phi],\, \phi\big] = \delta_{ij}\, \Delta_i^{\eminus1} - X_{ij}\,, \qquad \eta= (\Phi,\, \phi ),
    \end{equation}
where again $ \Delta^{\eminus 1} $ denotes the kinetic operator and $ X_{ij} $ are interaction terms. 
The master formula for the one-loop matching in terms of these objects is  
    \begin{align} \label{eq:Master}
    S_\EFT^{(\ell= 1)}= \frac{i}{2} \,\mathrm{STr}\,\ln\,\Delta^{\eminus1} \Big|_\mathrm{hard}- \frac{i}{2} \sum_{k=1}^\infty \frac{1}{k}\,\mathrm{STr}\big[ (\Delta X)^k\big] \Big|_\mathrm{hard}\, .
    \end{align}
Here, \textit{hard} indicates that loop integrands are expanded around loop momenta $ q\sim \Lambda $, following the method of expansion by regions~\cite{Beneke:1997zp,Jantzen:2011nz}. This form allows for a straightforward counting of light mass dimensions, and the resulting series can be truncated at the relevant order in the mass expansion.  

For the actual computation of Eq.~\eqref{eq:Master}, we follow the implementation outlined in Ref.~\cite{Fuentes-Martin:2020udw} based on the developments of Refs.~\cite{Fuentes-Martin:2016uol,Zhang:2016pja,Cohen:2020fcu}. The procedure allows for simultaneous treatment of all particle spins and mixed heavy and light states in the loop. Furthermore, the traces can be evaluated in a manifestly gauge covariant manner using the Covariant Derivative Expansion~(CDE)~\cite{Gaillard:1985uh,Chan:1986jq,Cheyette:1987qz}. Altogether, the method allows for a very algorithmic and efficient approach to evaluating all loop contributions simultaneously. Moreover, it remains possible to pinpoint specific contributions based on the fields propagating in the loops by targeting specific supertraces.

\subsection{Simplifications}
Properly simplifying the output Lagrangian is a challenge related to the long-standing problem of finding a basis for the higher-dimensional operators of an EFT. We distinguish between simplification with exact identities (integration-by-parts and group identities, and commutation relations), taking the Lagrangian to the Green's basis, and using field redefinitions to produce a simplified Lagrangian with on-shell equivalence. The exact simplification relates operators linearly, and can be applied to individual operators as well as the full EFT Lagrangian. On the other hand, field redefinitions work non-linearly and make sense only when acting on the EFT Lagrangian as a whole.

\subsubsection{Green's basis}
To reduce EFT Lagrangians to a Green's basis, we use methods from linear algebra, as this allows for efficient and robust simplifications. One can think of $ \L_\EFT $ as an element in the vector space $ O $ equipped with a basis $ \{\mathcal{O}_a \} $ consisting of all operators in the absence of any exact identities. That is, the elements of this basis span the complete set of gauge and Lorentz-invariant monomials of the fields, their covariant derivatives, and CG coefficients (including Dirac matrices). This vector space is redundant once the exact identities, relating the basis operators, are accounted for.  Each identity relation can be represented as a vector that is equivalent to $ 0 $. Together, the identity vectors span a subspace $ I \subseteq O $, and we can identify the coset $ O/I $ with the set of Green's basis Lagrangians. Simplifying $ \L_\EFT $ then comes down to finding a convenient basis for $ O/I $ and determining the representative element of the equivalence class $ [\L_\EFT] $ defined by the coset.

To arrive at the Green's basis in \matchete, we employ the following strategy: For all basis elements (operators) we encounter, we generate the complete set of possible identities using integration by parts, Jacobi identities, commutation of covariant derivatives, and gamma matrix identities (such as $ \gamma_\mu \gamma_\nu = g_{\mu\nu} - i \sigma_{\mu\nu} $). Denoting the vectors in $ O $ corresponding to the resulting identities by $ \mathcal{I}_n $, it follows that $ I = \mathrm{span} \big(\{ \mathcal{I}_n\} \big) $. The operator basis of $ O $ allows for the decomposition 
	\begin{equation}
	\mathcal{I}_n = \sum_a M_{na} \mathcal{O}_a,
	\end{equation}
which in turn defines a matrix $ M $ with the coordinate vectors of the identities as its rows. With standard methods, $ M $ is brought to reduced row echelon form $  M' $, and we observe that the non-zero rows describe a basis (in coordinate space) for $ I $. Conveniently, the first ``$ 1 $'' in each non-zero row  of $ M' $ effectively picks out a set of redundant operators $ \{\mathcal{O}_r\}_{r \in R} $, which can be eliminated in the EFT Lagrangian. The rows of $ M' $, thus, describe a set of identities for the equivalence classes:  
	\begin{equation}
		\bigg[\mathcal{O}_r + \sum_{b\notin R} M'_{rb} \mathcal{O}_{b} \bigg] = [0], \qquad \forall r \in R.
	\end{equation}
Using these identities, all $ \{\mathcal{O}_r\}_{r\in R} $ can be eliminated from the representative element of $ [\L_\EFT] $, that is, the Green's basis Lagrangian. By absorption of gauge couplings into gauge fields, all entries in $ M $ are numbers, allowing for efficient matrix manipulations of $ M $. The main challenge to implementing the simplification procedure described above is that of identifying identical operators based on their internal representation. To this end, \matchete relies heavily on pattern matching to identify, e.g., different labeling of the dummy indices, permutations of indices on symmetric tensors, and orderings of terms in products. By choosing an ordering of the basis $ \{\mathcal{O}_a \} $, it is possible to dictate a preference as to what operators are considered redundant, that is, in $ \{ \mathcal{O}_r\}_{r \in R} $. While the choice is somewhat arbitrary, we can ensure that the maximal number of operators that can be removed with field redefinitions are kept in the basis. Additional requirements are enforced to ensure that the output Lagrangian is manifestly Hermitian.

The main limitation of our current approach is the need to hard-code all possible identity types in \matchete. However, additional identities can be added in a modular manner. The initial version notably does not include Fierz identities, as the proper handling of these necessitate the evaluation of evanescent contributions~\cite{Fuentes-Martin:2022vvu}. This is something we expect to address in future updates. In any event, the lack of implementation of identities does not result in an invalid result from the simplification method, merely a non-minimal operator basis for $ \L_\EFT $ as the full identity space $ I $ is not found.

\subsubsection{Field redefinitions} \label{sec:field_redef}
\begin{table}[t]
\renewcommand{\arraystretch}{1.2}
\centering
\begin{tabular}{|c|c|c|}
\hline
\cellcolor{Gray!25} Field Type & \cellcolor{Gray!25} Objects & \cellcolor{Gray!25} Definition \\
\hline
Scalar $\varphi$ & $\eom{\varphi}$ & $ D_\mu D^\mu \varphi $ \\ \hline
\multirow{2}{*}{Dirac fermion $\psi$} & $\eom{\psi}$ & $\gamma_\mu D^\mu \psi$ \\
& $\eom{\bar \psi}$ & $(D^\mu \bar \psi) \gamma_\mu$ \\ \hline
\multirow{2}{*}{Majorana fermion $\eta$} & $\eom{\eta}$ & $\gamma_\mu D^\mu \psi$ \\
& $\eom{\eta^T}$ & $ (D^\mu \eta^T)\gamma_\mu^T$ \\ \hline
Vector $A$ & $\eom{A^\nu}$ & $D_\mu A^{\mu\nu}$ \\ \hline
\end{tabular}
\caption{Definitions of the operator $\eom{\psi}$ for the various field types. In the scalar and vector cases, operators acting on a complex-conjugated field follow straightforwardly from replacing the field with its complex conjugate in the definition. In the last line, $A^{\mu\nu}$ denotes the usual field-strength tensor associated with the vector field $A$.}
\label{tab:eomOperators}
\end{table}

After the simplifications outlined in the previous section have been performed, we are left with a Lagrangian that contains redundant operators that can be removed by field redefinitions. To classify these operators, we first define for each field type the object $\eom{\psi}$ corresponding to the kinetic piece of the field EOM and whose definition is given in Table~\ref{tab:eomOperators}. Operators with at least one occurrence of $\eom{\psi}$ can be removed from the Lagrangian employing a field redefinition of the field $\psi$. For operators at the highest power (that is, dimension-six operators when one is working up to dimensions six), such field redefinitions are equivalent to replacing the field EOM at leading power in the corresponding operator. In the presence of effective operators of different power-counting orders, this procedure misses some of the higher-order terms in power-counting and yields an incorrect result~\cite{Chisholm:1961tha, Kamefuchi:1961sb,Divakaran:1963yxz,Kallosh:1972ap,Salam:1970fso,Ball:1993zy,Arzt:1993gz,Criado:2018sdb}. We, therefore, employ field redefinitions for the sake of generality. 

The general procedure is as follows: First, one identifies all instances of~$\eom{\psi}$ for all fields~$\psi$ appearing in the Lagrangian. Then one reads off the overall coefficient $\eom{\psi}$ and performs a field redefinition by these coefficients. For illustration, consider a real scalar Lagrangian of the form
\begin{align}
\begin{aligned}
    \mathcal L &=  \frac{1}{2}(D_\mu\varphi)(D^\mu\varphi) - \frac{1}{2}m^2\varphi^2 + \frac{c}{\Lambda^2}\varphi^3D_\mu D^\mu\varphi \\
    &= \frac{1}{2}(D_\mu\varphi)(D^\mu\varphi) - \frac{1}{2}m^2\varphi^2 + \frac{c}{\Lambda^2}\varphi^3 \eom{\varphi}\,.
\end{aligned}
\end{align}
The field redefinition $\varphi \to \varphi + \frac{c}{\Lambda^2}\varphi^3$ removes the redundant operator when inserted into the kinetic term and produces a quartic operator when inserted into the mass term:
\begin{align}
    \mathcal L\to\mathcal L' = \frac{1}{2}(D_\mu\varphi)(D^\mu\varphi) - \frac{1}{2}m^2\varphi^2 - c \frac{m^2}{\Lambda^2}\varphi^4\,.
\end{align}
For complex fields as well as Majorana fermions, one reads off the coefficients of the conjugated EOM objects and averages over them. The complete list of field redefinitions for each field type is given in Table~\ref{tab:fieldRedefs}.

\begin{table}[t]
\renewcommand{\arraystretch}{1.2}
\centering
\begin{tabular}{|c|c|c|}
\hline
\cellcolor{Gray!25} Field Type & \cellcolor{Gray!25} Redundant operators & \cellcolor{Gray!25} Field redefinition \\ \hline
Real scalar $\varphi$ & $\chi \eom{\varphi}$ & $\varphi \to \varphi + \chi$ \\ 
Complex scalar $\phi$ & $\chi \eom{\phi} + \eom{\phi^\dagger}\Delta$ & $\phi \to \phi + \frac{1}{2}(\chi^\dagger + \Delta)$ \\ \hline
Majorana fermion $\eta$ & $\chi \eom{\eta} + \eom{\eta^T}\Delta$ & $\eta\to\eta + i C(\chi^T-\Delta)$ \\
Dirac fermion $\psi$ & $\chi \eom{\psi} +\eom{\bar\psi}\Delta$ & $\psi\to\psi - \frac{i}{2}(\bar\chi + \Delta)$ \\ \hline
Real vector field $A$ & $\eom{A_\mu}\chi^\mu$ & $A_\mu \to A_\mu - \chi_\mu$ \\
Complex vector field $A$ & $\eom{A_\mu}\chi^\mu + \eom{A^\dagger_\mu}\Delta^\mu$ & $A_\mu \to A_\mu -\frac{1}{2}( \chi_\mu^\dagger + \Delta_\mu) $ \\ \hline
\end{tabular}
\caption{Field redefinitions needed to remove redundant operators involving a given field type. See Table~\ref{tab:eomOperators} for the definitions of $\eom{\psi}$.}
\label{tab:fieldRedefs}
\end{table}

When eliminating a redundant operator with $ \eom{\psi} $, the field redefinitions will generate contributions only at higher order in the EFT power counting or at the same dimension but with fewer derivatives. 
In practice, therefore, one proceeds in an iterative fashion, seeking out redundant operators at the lowest order in the EFT counting and removing them by field redefinitions. The lowest-order operators that can appear here are kinetic-mixing terms at dimension four. At higher powers, the procedure needs to be repeated since operators may contain more than one $\eom{\psi}$ object, and the redefinition removes only one occurrence. Once all redundant terms are removed at a given order in power-counting, the procedure is repeated at the next order, until no more redundant operators remain.

Special care needs to be taken in the case of Abelian gauge fields. Since removing kinetic-mixing terms between gauge fields amounts to complicated redefinitions of the charges under the associated gauge groups, we choose to keep them explicit.\footnote{Kinetic-mixing terms are ubiquitous in BSM models with new $ \U(1) $ symmetries, which can mix with the hypercharge field. They are typically kept explicit in the Lagrangian until symmetry breaking.} Hence, in the presence of kinetic mixing, field redefinitions should be modified as follows: Consider a set of vector fields $A_\mu^i$ that exhibit kinetic mixing parameterized by a mixing matrix $Z$ and a redundant operator involving the vector fields:
    \begin{align}
    \mathcal{L} 
    = -\frac{1}{4} A^i_{\mu\nu}Z_{ij} A^{j,\mu\nu} + \chi_i^\nu D^\mu A^i_{\mu\nu} 
    \equiv -\frac{1}{4} A^i_{\mu\nu} Z_{ij} A^{j,\mu\nu} + \chi_i^\nu\, \eom{A^i_\nu}\,.
    \end{align}
The appropriate field redefinition is $A^i_\mu = A^i_\mu - (Z^{\eminus 1}\cdot \chi_\mu)^i$. As long as the deviation of $Z$ from identity is perturbative (in either loop counting or EFT power-counting), its inverse can be easily obtained.

An important subtlety arises from matching corrections to the couplings of operators of mass-dimension lower than four. That is, mass terms and cubic scalar interactions. For example---as is famously the case with the Higgs boson in the SM---if heavy degrees of freedom coupled to light scalars are integrated out at loop level, the latter receive mass corrections proportional to the hard scale. This upsets the power-counting of the effective theory, since a mass term for a light scalar of the form
\begin{align}
    \delta\mathcal L = -\frac{c_1}{2}\,\Lambda^2\,\varphi^2\,,
\end{align}
is formally of dimension two in the EFT counting even if $ c_1 $ is loop suppressed. In this case, \matchete introduces an effective coupling $m_{\varphi,\mathrm{eff}}^2$,
\begin{align}
 - \frac{1}{2}\left( m_\varphi^2 + c_1 \Lambda^2 \right) \varphi^2 \to -\frac{1}{2} m_{\varphi,\mathrm{eff}}^2 \, \varphi^2\,,
\end{align}
that is treated as EFT dimension two, such that the term is of dimension four again. In doing so, the program assumes a (fine-tuned) cancellation between the tree-level mass and the loop correction when the power enhancement from $\Lambda^2$ is large enough to overcome the loop suppression from $c_1$.

\subsection{Conventions}
In this section we clearly state the overarching conventions used in \matchete to prevent unnecessary confusion on the matter. For the metric we use the ``mostly-minus" signature: $g_{\mu \nu} = \diag(+1, -1, -1, -1) $. Meanwhile, we take the antisymmetric Levi--Civita tensor to satisfy $ \varepsilon^{0 1 2 3} = - \varepsilon_{0 1 2 3} = +1 $ while the chiral spinor projectors are $P_\LL = \tfrac{1}{2}(1 - \gamma_5)$ and $P_\RR = \tfrac{1}{2}(1 + \gamma_5) $. The covariant derivatives of the gauge groups are automatically generated and used throughout the package. They are defined by $D_\mu = \partial_\mu - i g T^a A^a_\mu$ (note the sign) for non-Abelian groups with gauge field $ A_\mu $, with $ g $ being the coupling and $T^a$ the Hermitian generators, which normalize as $ \tr[T^a T^b] = \tfrac{1}{2} \delta^{ab} $ for fundamental representations. For the Abelian gauge groups, $ T^a $ is replaced by the charge. 

All computations are performed in dimensional regularization (DR) with spacetime dimension $d = 4 - 2 \epsilon $. The renormalization scheme is \msbar in line with most BSM computations. The treatment of $ \gamma_5 $ is a point of contention in DR and fraught with potential errors. In this initial release, we employ naive dimensional regularization (NDR). Namely, we use the anticommuting $ \gamma_5 $ and impose the four-dimensional identity          
    \begin{equation} \label{eq:SNDR_pescription}
    \tr[ \gamma^\mu \gamma^\nu \gamma^\rho \gamma^\sigma \gamma_5 ] = - 4 i \varepsilon^{\mu \nu \rho \sigma}. 
    \end{equation}
Trace cyclicity is lost in this manner, but as long as the EFT computations follow the Dirac-trace reading point of the matching computation, all ambiguities cancel~\cite{Fuentes-Martin:2020udw}. The source of the few ambiguities in the matching stem from IR divergences in loops with heavy and light fermions. In these cases, \matchete reading points can be inferred, since it will always read these supertraces starting with a heavy-fermion propagator. This is not a particularly elegant solution, and we plan to explore other approaches for handling $ \gamma_5 $ in future updates.

\section{Using Matchete} \label{sec:examples}

The \matchete package is free software under the terms of the GNU General Public License~v3.0 and is publicly available in the GitLab repository
\begin{center}
\href{https://gitlab.com/matchete/matchete}{https://gitlab.com/matchete/matchete}
\end{center}
The package can be installed in one of two ways:
\begin{enumerate}[i)]
    \item \textit{Automatic installation}: The simplest way to download and install \matchete is to run the following command in a \pkg{Mathematica} notebook:  
\mmaSet{leftmargin=15pt}
\begin{mmaCell}{Input}
  Import["https://gitlab.com/matchete/matchete/-/raw/master/install.m"]
\end{mmaCell}
    This will download and install \matchete in the \textit{Applications} folder of \pkg{Mathematica}'s base directory.
    
    \item \textit{Manual installation}: The user can also manually download the package from the GitLab repository. In this case, the user has to specify the location of the downloaded package with\footnote{We recommend placing the \matchete folder in the \textit{Applications} folder of \pkg{Mathematica}'s base directory. Then the location does not need be specified before loading the package.}
\mmaSet{index=1}
\mmaSet{leftmargin=3em}
\begin{mmaCell}{Input}
  PrependTo[\$Path,"directory"];
\end{mmaCell}
    where \texttt{\color{DarkGray}directory} is the path to the \matchete folder.
\end{enumerate}
Once installed, the user can load \matchete in a fresh \pkg{Mathematica} kernel by running:
\begin{mmaCell}{Input}
  << \mmaDef{Matchete\(\,\grave\,\)}
\end{mmaCell}
The user can check for updates and install them (when available) by simply running the \texttt{CheckForUpdate[]} command in a \pkg{Mathematica} notebook.

Once \matchete is installed and loaded, the user can start implementing models and matching to their EFTs with the routines provided by the package. Below, we demonstrate the usage of the tool with illustrative examples.

\subsection{Vector-like fermion toy-model}
To illustrate the use of \matchete with a simple but comprehensive example, we consider a variation of the toy model of Ref.~\cite{Fuentes-Martin:2020udw} with a $\U(1)$~gauge symmetry, two charged vector-like fermions $\psi$ and $\Psi$, and a real scalar singlet~$\phi$. The Lagrangian is given by
\begin{align}
\mathcal{L}&= -\frac{1}{4}F_{\mu\nu}F^{\mu\nu}+\frac{1}{2}(\partial_\mu \phi)^2 - \frac{1}{2}m_\phi^2 \phi^2 +\bar \psi\,i \slashed D\, \psi + \bar \Psi(i\slashed{D}-M)\Psi-\left(y\,\bar \psi_L\,\phi\, \Psi_R+\mathrm{h.c.}\right),
\end{align}
where $D_\mu\psi=\partial_\mu\psi-ie\, A_\mu\psi$ (and similarly for $\Psi$).
We take $\psi$ to be massless, $\phi$ to have a light mass $m_\phi$, and $\Psi$ to have a heavy mass $M$. The low-energy EFT, describing physics at energies much lower than $ M $, is obtained by integrating out~$\Psi$. We proceed to show how the matching is performed in \matchete.\footnote{A preliminary implementation of this model in \matchete was also presented in \cite{Wilsch:2022kfv}.}

As a first step, the user has to define all (gauge)~symmetries of the theory. We define the $\U(1)$ symmetry of the present example, labeled~\mmaInlineCell[]{Input}{U1e}, by
\begin{mmaCell}{Input}
  DefineGaugeGroup[U1e, U1, e, A]
\end{mmaCell}
which initializes a gauge coupling~\mmaInlineCell[]{Input}{e} and the corresponding field-strength tensor, labeled~\mmaInlineCell[]{Input}{A}.
Next, all matter fields are defined: 
\begin{mmaCell}{Input}
   DefineField[\mmaUnd{\(\Psi\)}, Fermion, Charges -> \{U1e[1]\}, Mass -> \{Heavy, M\}]\newline DefineField[\mmaUnd{\(\psi\)}, Fermion, Charges -> \{U1e[1]\}, Mass -> 0]\newline DefineField[\mmaUnd{\(\phi\)}, Scalar, Mass -> Light, SelfConjugate -> True]
\end{mmaCell}
where we assign charges of $ +1 $ to both fermions under the $\U(1)$ gauge group, declare the field~\mmaInlineCell[]{Input}{\mmaUnd{\(\Psi\)}} as heavy (for matching purposes) with mass label~\mmaInlineCell[]{Input}{\(M\)}, set~\mmaInlineCell[]{Input}{\mmaUnd{\(\psi\)}} as massless, and set~\mmaInlineCell[]{Input}{\mmaUnd{\(\phi\)}} to be a real field with light mass, automatically generating the mass label \mmaInlineCell[]{Input}{\mmaUnd{\(m\phi\)}}.  Finally, we have to define the Yukawa coupling~\mmaInlineCell[]{Input}{y}
\begin{mmaCell}{Input}
  DefineCoupling[y]
\end{mmaCell}
which by default is understood as a complex parameter that does not influence the EFT power counting (i.e. \mmaInlineCell[]{Input}{EFTOrder -> 0}).

After all symmetries, fields, and couplings are defined, the Lagrangian of the free theory can be automatically generated with the~\mmaInlineCell[]{Input}{FreeLag} routine. The interactions are manually added to obtain the UV Lagrangian
\begin{mmaCell}{Input}
  LUV = FreeLag[] - PlusHc[\mmaDef{y}[]\,\(\phi\)[]\,Bar[\(\psi\)[]]**PR**\(\Psi\)[]];
\end{mmaCell}
where the~\mmaInlineCell[]{Input}{PlusHc} routine automatically adds the Hermitian conjugate of its argument.
With the~\mmaInlineCell[]{Input}{NiceForm} formatting, we can then verify that the Lagrangian does in fact agree with our expectations:
\begin{mmaCell}{Input}
  \mmaDef{LUV} //NiceForm
\end{mmaCell}
\begin{mmaCell}{Output}
  - \(\frac{1}{4}\) \!\!\mmaSup{\mmaSup{A}{\(\mu\nu\)}}{2} + \(\frac{1}{2}\) \!\!\mmaSup{(\mmaSub{D}{\(\mu\)}\(\phi\))}{2} - \(\frac{1}{2}\)\mmaSup{\(m\phi\)}{2} \!\!\mmaSup{\(\phi\)}{2} + i(\(\overline{\psi}\) \!\!\(\cdot\)\!\! \mmaSub{\(\gamma\)}{\(\mu\)} \!\!\(\cdot\)\!\! \mmaSub{D}{\(\mu\)}\(\psi\)) + i(\(\overline{\Psi}\) \!\!\(\cdot\)\!\! \mmaSub{\(\gamma\)}{\(\mu\)} \!\!\(\!\cdot\)\!\! \mmaSub{D}{\(\mu\)}\(\Psi\)) - M(\(\overline{\Psi} \!\!\cdot\)\!\! \(\Psi\)) \newline\newline - y\(\phi\) \!\!(\(\overline{\psi}\) \!\!\(\cdot\)\!\! \mmaSub{P}{R} \!\!\(\cdot\)\!\! \(\Psi\)) - \(\overline{y}\phi\) \!\!(\(\overline{\Psi}\cdot\)\!\! \mmaSub{P}{R} \!\!\(\cdot\)\!\! \(\psi\))
\end{mmaCell}    
Next, we integrate out the heavy fermion~\mmaInlineCell[]{Input}{\(\Psi\)} with the~\mmaInlineCell[]{Input}{Match} routine:
\begin{mmaCell}{Input}
  LEFT = Match[\mmaDef{LUV}, LoopOrder -> 1, EFTOrder -> 6];
\end{mmaCell}
where the option~\mmaInlineCell[]{Input}{EFTOrder -> 6} prescribes the EFT expansion is terminated at dimension-six operators, and~\mmaInlineCell[]{Input}{LoopOrder -> 1} indicates that the matching is performed at one-loop order. The resulting EFT Lagrangian~\mmaInlineCell[]{Input}{\mmaDef{LEFT}} is given in a redundant, unsimplified form. It can be simplified to an off-shell Green's basis by calling
\begin{mmaCell}{Input}
  LEFTOffShell = GreensSimplify[\mmaDef{LEFT}];
\end{mmaCell}
More commonly, we wish to also use field redefinitions to achieve an even more simplified EFT that still reproduces the same on-shell physics. Simplification to the on-shell basis is performed by the means of field redefinitions (see Sec.~\ref{sec:field_redef}) by calling the~\mmaInlineCell[]{Input}{EOMSimplify} routine:
\begin{mmaCell}{Input}
  LEFTOnShell = EOMSimplify[\mmaDef{LEFT}] /.\mmaSup{\(\epsilon\)}{-1} -> 0 //NiceForm
\end{mmaCell}
\begin{mmaCell}{Output}
  \Big(-\mmaFrac{1}{4} - \mmaFrac{1}{3}\,\(\hbar\)\,\mmaSup{e}{2}\,Log\Big[\mmaFrac{\mmaSup{\(\overline{\mu}\)}{2}}{\mmaSup{M}{2}}\Big]\Big)\,\mmaSup{\mmaSup{A}{\(\mu\nu\)}}{2} + \mmaFrac{1}{2}\,\mmaSup{(\mmaSub{D}{\(\mu\)}\(\phi\))}{2} + i\,(\(\overline{\psi}\,\cdot\)\,\mmaSub{\(\gamma\)}{\(\mu\)}\,\(\cdot\)\,\mmaSub{D}{\(\mu\)}\(\psi\))\newline\newline+\Bigg(c\(\phi\phi\) + \mmaFrac{1}{3}\,\(\hbar\)\,\(\overline{y}\)y\,c\(\phi\phi\)\,\mmaFrac{1}{\mmaSup{M}{2}}\,\Big(4\,c\(\phi\phi\) - 3\mmaSup{M}{2}\Big(1 + 2\,Log\Big[\mmaFrac{\mmaSup{\(\overline{\mu}\)}{2}}{\mmaSup{M}{2}}\Big]\Big)\Big)\Bigg)\mmaSup{\(\phi\)}{2}\newline\newline+\mmaFrac{1}{9}\,\(\hbar\)\,\mmaSup{\(\overline{y}\)}{2}\,\mmaSup{y}{2}\,\mmaFrac{1}{\mmaSup{M}{2}}\,\Big(13\,c\(\phi\phi\) - 9\,\mmaSup{M}{2}\,Log\Big[\mmaFrac{\mmaSup{\(\overline{\mu}\)}{2}}{\mmaSup{M}{2}}\Big]\Big)\,\mmaSup{\(\phi\)}{4} + \mmaFrac{1}{3}\,\(\hbar\)\,\mmaSup{\(\overline{y}\)}{3}\,\mmaSup{y}{3}\,\mmaFrac{1}{\mmaSup{M}{2}}\,\mmaSup{\(\phi\)}{6}\newline\newline+\mmaFrac{1}{3}\,\(\hbar\)\,\mmaFrac{\(\overline{y}\)y\,\mmaSup{e}{2}}{\mmaSup{M}{2}}\,\mmaSup{\(\phi\)}{2}\,\mmaSup{\mmaSup{A}{\(\mu\nu\)}}{2} - \mmaFrac{2}{15}\,\(\hbar\)\,\mmaFrac{\mmaSup{e}{4}}{\mmaSup{M}{2}}\,\mmaSup{(\(\overline{\psi}\,\cdot\)\,\mmaSub{\(\gamma\)}{\(\mu\)}\,\(\cdot\,\psi\))}{2} + \mmaFrac{7}{36}\,\(\hbar\)\,\mmaFrac{\(\overline{y}\)y\,\mmaSup{e}{2}}{\mmaSup{M}{2}}\,(\(\overline{\psi}\,\cdot\)\,\mmaSub{\(\gamma\)}{\(\mu\)}\,\(\cdot\,\psi\))(\(\overline{\psi}\,\cdot\)\,\mmaSub{\(\gamma\)}{\(\mu\)}\mmaSub{P}{L}\,\(\cdot\,\psi\))
\end{mmaCell}
where we set all the poles to zero, assuming that both the UV Lagrangian~\mmaInlineCell[]{Output}{\mmaDef{LUV}} and the EFT Lagrangian~\mmaInlineCell[]{Output}{\mmaDef{LEFTonShell}} are properly renormalized in the \msbar-scheme.
In a slight abuse of notation, $\hbar$ is used in the output to denote the loop factor and ensure consistent truncation of the loop expansion, i.e. for one-loop computations \matchete sets~$\hbar^2=0$. For numerical values, one simply needs to replace $\hbar\to {1}/{(16\pi^2)}$. We observe that there are no redundant operators left in this EFT Lagrangian. The simplified output has canonically normalized kinetic terms for the matter fields, leaving only the non-trivial factor on the gauge kinetic term in lieu of a coupling correction.  

The coupling~\mmaInlineCell[]{Output}{c\(\phi\phi\)} is automatically introduced to account for the hard scale contribution to the mass correction of the scalar field, as described in Sec.~\ref{sec:field_redef}. The user is notified when such replacements happen and can retrieve the definitions of the effective couplings in the resulting Lagrangian using the \texttt{PrintEffectiveCouplings} command:
\begin{mmaCell}{Input}
  \mmaDef{PrintEffectiveCouplings}[\mmaDef{LEFTOnShell}] 
\end{mmaCell}
\begin{mmaCell}{Output}
  c\(\phi\phi\) = -\mmaFrac{1}{2}\,\mmaSup{m\(\phi\)}{2} - 2\,\(\hbar\)\,\(\overline{y}\)y\,\mmaSup{M}{2} - 2\,\(\hbar\)\,\(\overline{y}\)y\,\mmaSup{M}{2}\,Log\Big[\mmaFrac{\mmaSup{\(\overline{\mu}\)}{2}}{\mmaSup{M}{2}}\Big]
\end{mmaCell}

\noindent If desired, the effective couplings can be replaced by their definitions in terms of the original input couplings using the \texttt{ReplaceEffectiveCouplings} command.

\subsection{Real singlet scalar BSM extension}\label{sec:singletscalar}
\mmaSet{index=1}
Our first BSM example is the venerable singlet extension of the SM previously matched in Refs.~\cite{Jiang:2018pbd,Haisch:2020ahr}. A real, heavy scalar field $ \phi $, which is a singlet under the SM gauge group is added to the SM. The resulting Lagrangian for this UV model is 
    \begin{equation}
    \L = \L_\sscript{SM} + \tfrac{1}{2} (\partial_\mu \phi)^2 - \tfrac{1}{2} M^2 \phi - \frac{\mu}{3!} \phi^3 -\frac{\lambda_\phi}{4!} \phi^4 - A \phi |H|^2 - \frac{\kappa}{2} \phi^2 |H|^2 .
    \end{equation}
Assuming the mass of the scalar to be heavy compared to the electroweak scale, the singlet can be integrated out from the theory to arrive at the corresponding SMEFT Lagrangian. We have validated the full one-loop dimension-six result of this matching and obtained agreement with the calculation of Ref.~\cite{Haisch:2020ahr}.\footnote{Actually, agreement with this reference is only obtained for $\bar\mu=M^2$, as we find that the results provided there for the log-terms contain contributions that cannot be generated by matching. The log-terms have been partially cross-checked against the Greens' basis results in Ref.~\cite{Cohen:2020fcu}, finding agreement for the non-logarithmic contributions as well.} Here, we will show how this simple SM extension can be implemented in \matchete and how to select specific contributions from the matching computation. 

Since we are dealing with a SM extension, the task of inputting the model is simpler. The first step is to load the SM Lagrangian, which is already predefined in \matchete, by running\footnote{The complete list of available models (including the one of this example) can be checked by \texttt{GetModels[]} in a \pkg{Mathematica} notebook or by looking into the Models folder of the public release.}
\begin{mmaCell}{Input}
  LSM = LoadModel["SM", ModelParameters -> \{"\(\mu\)" -> mH, "\(\lambda\)" -> \mmaUnd{\(\lambda\)h}\}];
\end{mmaCell}
where we rename the Higgs mass parameter to \mmaInlineCell[]{Input}{mH} and the quartic Higgs coupling to \mmaInlineCell[]{Input}{\mmaUnd{\(\lambda\)h}}. 
This command defines all SM symmetries, couplings and fields, and saves the SM Lagrangian into the \mmaInlineCell[]{Input}{\mmaDef{LSM}} variable.
For completeness, we also provide the full SM definition in \matchete in Appendix~\ref{app:SM}. The implementation shown there agrees with the internal implementation that is loaded when using~\mmaInlineCell[]{Input}{LoadModel["SM"]}.

Next, we have to define the BSM field~\mmaInlineCell[]{Input}{\mmaUnd{\(\phi\)}} with mass~\mmaInlineCell[]{Input}{M} by
\begin{mmaCell}{Input}
  DefineField[\mmaUnd{\(\phi\)}, Scalar, SelfConjugate -> True, Mass -> \{Heavy, \mmaUnd{M}\}]
\end{mmaCell}
followed by the definition of all NP couplings:
\begin{mmaCell}{Input}
   DefineCoupling[\mmaUnd{A}, SelfConjugate -> True]\newline DefineCoupling[\mmaUnd{\(\kappa\)}, SelfConjugate -> True]\newline DefineCoupling[\mmaUnd{\(\mu\)}, SelfConjugate -> True]\newline DefineCoupling[\mmaUnd{\(\lambda\phi\)}, SelfConjugate -> True]
\end{mmaCell}
Using these definitions, none of the couplings above carry a light mass dimension, i.e., we have \mmaInlineCell[]{Input}{\mmaUnd{\(\mu\)}}\,$=\mathcal{O}(M)$ and~\mmaInlineCell[]{Input}{A}\,$=\mathcal{O}(M)$.
The Lagranigian of the full NP model can then be specified with
\begin{mmaCell}{Input}
  LUV = \mmaDef{LSM} + FreeLag[\(\phi\)] - \mmaFrac{1}{3!}\(\mu\)[]\,\mmaSup{\(\phi\)[]}{3} - \mmaFrac{1}{4!}\(\lambda\phi\)[]\,\mmaSup{\(\phi\)[]}{4}\newline      - \mmaDef{A}[]\,Bar[H[i]]H[i]\(\phi\)[] - \mmaFrac{1}{2}\(\kappa\)[]\,Bar[H[i]]H[i]\mmaSup{\(\phi\)[]}{2};
\end{mmaCell}

The matching to the SMEFT is again performed with the \mmaInlineCell[]{Input}{Match} routine. For the tree-level matching, we find
\begin{mmaCell}{Input}
   LEFT0 = Match[\mmaDef{LUV}, LoopOrder -> 0, EFTOrder -> 6];\newline GreensSimplify[LEFT0 - \mmaDef{LSM}] //HcSimplify //NiceForm
\end{mmaCell}
\begin{mmaCell}{Output}
   \mmaFrac{1}{2}\,\mmaSup{A}{2}\,\mmaFrac{1}{\mmaSup{M}{2}}\,\mmaSub{\(\overline{H}\)}{i}\mmaSub{\(\overline{H}\)}{j}\mmaSup{H}{i}\mmaSup{H}{j} + \mmaFrac{1}{6}\,\mmaSup{A}{2}\,\mmaFrac{1}{\mmaSup{M}{6}}\,\big(-3\(\kappa\)\,\mmaSup{M}{2} + A\,\(\mu\)\big)\,\mmaSub{\(\overline{H}\)}{i}\mmaSub{\(\overline{H}\)}{j}\mmaSub{\(\overline{H}\)}{k}\mmaSup{H}{i}\mmaSup{H}{j}\mmaSup{H}{k}\newline\newline -\mmaSup{A}{2}\,\mmaFrac{1}{\mmaSup{M}{4}}\,\mmaSub{\(\overline{H}\)}{i}\mmaSub{D}{\(\mu\)}\mmaSub{\(\overline{H}\)}{j}\mmaSup{H}{i}\mmaSub{D}{\(\mu\)}\mmaSup{H}{j} + \Big(-\mmaFrac{1}{2}\,\mmaSup{A}{2}\,\mmaFrac{1}{\mmaSup{M}{4}}\,\mmaSub{\(\overline{H}\)}{i}\mmaSub{\(\overline{H}\)}{j}\mmaSup{H}{i}\mmaSup{D}{2}\mmaSup{H}{j} + H.c.\Big)
\end{mmaCell}
where we only print the NP contributions in the EFT after applying off-shell operator simplification, such as integration-by-parts identities. The first operator is a modification of the Higgs quartic coupling, the second is $Q_{H}$ in the Warsaw basis (defined in~\cite{Grzadkowski:2010es}), the third can be exchanged for the operator $Q_{H\Box}$ of the Warsaw basis,\footnote{\matchete automatically applies the product rule for derivatives. Therefore, it is not possible to directly obtain $Q_{H\Box}$ in the matching result. The issues related to this basis mismatch are discussed below.} and the last term can be removed by applying appropriate field redefinitions or the Higgs EOM. This last simplification step can be performed by applying~\mmaInlineCell[]{Input}{EOMSimplify[\mmaDef{LEFT0}]}. For brevity, we do not show the result here, but it can be found in the example notebook \texttt{Examples/Singlet\_Scalar\_Extension.nb}, included in the public release of \matchete. This notebook contains the full matching of this model at one loop as well as the comparison to the results presented in Ref.~\cite{Haisch:2020ahr}.

The one-loop matching and the full simplification of the resulting EFT Lagrangian is performed similarly:
\begin{mmaCell}{Input}
   LEFT = Match[\mmaDef{LUV}, LoopOrder -> 1, EFTOrder -> 6];\newline LEFTOnShell = LEFT //EOMSimplify;
\end{mmaCell}
Again, the resulting Lagrangian is too long to show here, but it can be found in the example notebook. In the following, we demonstrate how to extract a particular contribution from the EFT Lagrangian, using the \mmaInlineCell[]{Input}{SelectOperatorClass} routine. As an example, we extract the fully leptonic four-fermion operator
\begin{mmaCell}{Input}
  SelectOperatorClass[\mmaDef{LEFTOnShell},\{Bar[\mmaDef{l}],\mmaDef{e},Bar[\mmaDef{e}],l\},0] //NiceForm
\end{mmaCell}
\begin{mmaCell}{Output}
  \mmaFrac{1}{6} \!\!\(\hbar\)\,\mmaSup{\(\overline{Ye}\)}{rs}\mmaSub{Ye}{tp}\,\mmaSup{A}{2}\mmaFrac{1}{\mmaSup{M}{4}}\big(\mmaSup{\(\overline{e}\)}{s}\,\(\cdot\)\,\mmaSub{P}{L}\,\(\cdot\)\,\mmaSup{l}{ir}\big)\big(\mmaSubSup{\(\overline{l}\)}{i}{t}\,\(\cdot\)\,\mmaSub{P}{R}\,\(\cdot\)\,\mmaSup{e}{p}\big)
\end{mmaCell}
where the second argument specifies the field content of the operator(s) to be extracted, and the last argument gives the number of derivatives.
The result shown above is not in the Warsaw basis, since the current version \matchete is not applying Fierz identities. Manually, using the identity $(\overline{e}^s\ell^r)(\overline{\ell}^t e^p)=-\frac{1}{2}(\overline{\ell}^t \gamma_\mu \ell^r)(\overline{e}^s \gamma^\mu e^p)$ gives the desired result for~$Q_{\ell e}$ in the Warsaw basis. Similarly, we extract the $Q_{HW}$ operator:
\begin{mmaCell}{Input}
  SelectOperatorClass[\mmaDef{LEFTOnShell},\{Bar[\mmaDef{H}],\mmaDef{H},\mmaDef{W},\mmaDef{W}\},0] //NiceForm
\end{mmaCell}
\begin{mmaCell}{Output}
  \mmaFrac{1}{12}\,\(\hbar\)\,\mmaSup{gL}{2}\,\mmaSup{A}{2}\,\mmaFrac{1}{\mmaSup{M}{4}}\,\mmaSub{\(\overline{H}\)}{i}\,\mmaSup{H}{i}\mmaSup{\mmaSup{W}{\(\mu\nu\)I}}{2}
\end{mmaCell}

In general, results obtained with the \mmaInlineCell[]{Input}{SelectOperatorClass} routine do not coincide with the matching conditions for the Warsaw basis. This is because of the $Q_{H\Box}=(H^\dagger H)\Box(H^\dagger H)$ operator being replaced in favor of the operator $Q_{HD}^{\,\prime}=(H^\dagger H)[(D_\mu H^\dagger) (D^\mu H)]$ in \matchete. These operators are related by integration by parts, but their difference is an operator that can be removed by applying Higgs field redefinitions. Therefore, the choice between $Q_{H\Box}$ or $Q_{HD}^{\,\prime}$ affects the matching conditions for a wide set of different operator classes. The examples shown here are, however, not affected by this. In the example notebook, we show how to manually match the results provided by \matchete to the Warsaw basis.

\subsection{Vector-like lepton BSM extension}\label{sec:VLLexample}
\mmaSet{index=2}
As our final example, we consider a vector-like lepton extension of the SM with the same quantum numbers as the SM lepton singlet, namely $E\sim(\rep{1},\rep{1})_{\eminus 1}$. The Lagrangian for this model is given by
\begin{align}\label{eq:VLL_Lag}
\mathcal{L}=\mathcal{L}_\sscript{SM}+\overline{E}(i\slashed{D}-M_E) E-\left(y_E^p\,\bar\ell^p H\,E_R\hc\right),
\end{align}
where $H$ is the SM Higgs, $\ell^p$ is the SM $\mathrm{SU}(2)_L$ lepton doublet, and the index~$p$ denotes SM flavor. The NP model parameters $M_E$ and $y_E^p$ are a real scalar and a complex flavor vector, respectively. The matching result of this model was already presented as an example for the \texttt{matchmakereft} matching tool~\cite{Carmona:2021xtq}, which uses a diagrammatic approach. We find full agreement with this result, hence providing essential validation for both implementations.

In what follows, we show how to input the Lagrangian~\eqref{eq:VLL_Lag} and illustrate the \matchete functions relevant to the cross check. As in the previous example, we are dealing with a SM extension and the first step is to load the SM Lagrangian, which is already predefined:
\begin{mmaCell}{Input}
  LSM = LoadModel["SM"];
\end{mmaCell}
This command defines all SM symmetries, couplings and fields, and saves the SM Lagrangian into the \mmaInlineCell[]{Input}{\mmaDef{LSM}} variable. The next step is to define the NP field\footnote{We write \texttt{EE} for the field label because \texttt{E} is reserved in \pkg{Mathematica} for the Euler number.}
\begin{mmaCell}{Input}
  DefineField[EE, Fermion, Charges -> \{U1Y[-1]\}, Mass -> \{Heavy, ME\}]
\end{mmaCell}
and Yukawa coupling
\begin{mmaCell}{Input}
  DefineCoupling[yE, Indices -> \{Flavor\}]
\end{mmaCell}
The complete UV Lagrangian is then entered as 
\begin{mmaCell}{Input}
  LUV = \mmaDef{LSM} \!+\! FreeLag[\mmaDef{EE}] \!-\! PlusHc[\mmaDef{yE}[p]\! Bar[\mmaDef{l}[i, p]]**PR**\mmaDef{EE}[] \!\mmaDef{H}[i]];
\end{mmaCell}
where \mmaInlineCell[]{Input}{i} and \mmaInlineCell[]{Input}{p} are used for $\mathrm{SU}(2)_L$ and flavor indices, respectively.
As in previous examples, this Lagrangian can easily be matched to its EFT with the \mmaInlineCell[]{Input}{Match} routine. For example, at tree level we have
\begin{mmaCell}{Input}
   LEFT0 = Match[\mmaDef{LUV}, LoopOrder -> 0, EFTOrder -> 6];\newline LEFT0 - \mmaDef{LSM} //NiceForm
\end{mmaCell}
\begin{mmaCell}{Output}
  i \!\!\mmaSup{\(\overline{yE}\)}{p}\mmaSup{yE}{r} \!\!\mmaFrac{1}{\mmaSup{ME}{2}} \!\!\mmaSub{D}{\(\mu\)}\mmaSub{\(\overline{H}\)}{i}\mmaSup{H}{j} \!\!(\mmaSubSup{\(\overline{l}\)}{j}{r} \!\!\(\cdot\) \!\!\mmaSub{\(\gamma\)}{\(\mu\)}\mmaSub{P}{L} \!\!\(\cdot\)\!\! \mmaSup{l}{ip}) \!+\! i \!\!\mmaSup{\(\overline{yE}\)}{p}\mmaSup{yE}{r} \!\!\mmaFrac{1}{\mmaSup{ME}{2}} \!\!\mmaSub{\(\overline{H}\)}{i}\mmaSup{H}{j} \!\!(\mmaSubSup{\(\overline{l}\)}{j}{r} \!\!\(\cdot\)\!\! \mmaSub{\(\gamma\)}{\(\mu\)}\mmaSub{P}{L} \!\(\cdot\)\! \mmaSub{D}{\(\mu\)}\mmaSup{l}{ip})
\end{mmaCell}
This result is not manifestly Hermitian but it can be recast into a manifestly Hermitian form using IBP identities via the \mmaInlineCell[]{Input}{GreensSimplify} routine:
\begin{mmaCell}{Input}
  \mmaDef{LEFT0} - \mmaDef{LSM} //GreensSimplify //NiceForm
\end{mmaCell}
\begin{mmaCell}{Output}
  \mmaFrac{i}{2} \!\!\mmaSup{\(\overline{yE}\)}{p}\mmaSup{yE}{r} \!\!\mmaFrac{1}{\mmaSup{ME}{2}} \!\!\Big(\mmaSub{D}{\(\mu\)}\mmaSub{\(\overline{H}\)}{i}\mmaSup{H}{j} \!\!(\mmaSubSup{\(\overline{l}\)}{j}{r} \!\!\(\cdot\) \!\!\mmaSub{\(\gamma\)}{\(\mu\)}\mmaSub{P}{L} \!\!\(\cdot\)\!\! \mmaSup{l}{ip}) \!-\!  \!\!\mmaSub{\(\overline{H}\)}{i}\mmaSub{D}{\(\mu\)}\mmaSup{H}{j} \!\!(\mmaSubSup{\(\overline{l}\)}{j}{r} \!\!\(\cdot\) \!\!\mmaSub{\(\gamma\)}{\(\mu\)}\mmaSub{P}{L} \!\!\(\cdot\)\!\! \mmaSup{l}{ip})\Big) \!+\!\newline\newline \mmaFrac{i}{2} \!\!\mmaSup{\(\overline{yE}\)}{p}\mmaSup{yE}{r} \!\!\mmaFrac{1}{\mmaSup{ME}{2}} \!\!\Big(\mmaSub{\(\overline{H}\)}{i}\mmaSup{H}{j} \!\!(\mmaSubSup{\(\overline{l}\)}{j}{r} \!\!\(\cdot\)\!\! \mmaSub{\(\gamma\)}{\(\mu\)}\mmaSub{P}{L} \!\(\cdot\)\! \mmaSub{D}{\(\mu\)}\mmaSup{l}{ip}) \!-\!  \!\!\mmaSub{\(\overline{H}\)}{i}\mmaSup{H}{j} \!\!(\mmaSubSup{\mmaSub{D}{\(\mu\)}\(\overline{l}\)}{j}{r} \!\!\(\cdot\)\!\! \mmaSub{\(\gamma\)}{\(\mu\)}\mmaSub{P}{L} \!\(\cdot\)\! \mmaSup{l}{ip})\Big)
\end{mmaCell}
Finally, this last term of the result above can be eliminated in favor of a Warsaw basis operator using field redefinitions (which at this order are equivalent to EOM identities). The field redefinitions are applied with the \mmaInlineCell[]{Input}{EOMSimplify} routine:
\begin{mmaCell}{Input}
   LEFT0OnShell = \mmaDef{LEFT0} //EOMSimplify;\newline LEFT0OnShell - \mmaDef{LSM} //HcSimplify //NiceForm
\end{mmaCell}
\begin{mmaCell}{Output}
  \Big(\mmaFrac{1}{2} \!\!\mmaSup{\(\overline{yE}\)}{s}\mmaSup{yE}{r}\mmaSup{Ye}{sp} \!\!\mmaFrac{1}{\mmaSup{ME}{2}} \!\!\mmaSub{\(\overline{H}\)}{i}\mmaSup{H}{i}\mmaSup{H}{j} \!\!(\mmaSubSup{\(\overline{l}\)}{j}{r} \!\!\(\cdot\) \!\mmaSub{P}{R} \!\!\(\cdot\)\! \mmaSup{e}{p}) \!\!-\!\! \mmaFrac{i}{2} \!\!\mmaSup{\(\overline{yE}\)}{p}\mmaSup{yE}{r} \!\!\mmaFrac{1}{\mmaSup{ME}{2}} \!\!\mmaSub{\(\overline{H}\)}{i}\mmaSub{D}{\(\mu\)}\mmaSup{H}{j} \!\!(\mmaSubSup{\(\overline{l}\)}{j}{r} \!\!\(\cdot\) \!\!\mmaSub{\(\gamma\)}{\(\mu\)}\mmaSub{P}{L} \!\!\(\cdot\)\!\! \mmaSup{l}{ip}) \!\!+\!\! H.c\Big)
\end{mmaCell}
A final group identity, $\delta_{ik}\delta_{jl}=\tfrac{1}{2}(\delta_{ij}\delta_{kl}+\tau^a_{ij}\tau^a_{kl})$, would be needed to recast the last term into elements of the Warsaw basis. At present, the automated reduction to the Warsaw basis is not implemented in \matchete and this and other identities need to be applied manually. In this example we were careful to apply \mmaInlineCell[]{Input}{EOMSimplify} to the full EFT Lagrangian. Unlike the exact identities used by \mmaInlineCell[]{Input}{GreensSimplify}, \emph{field redefinitions cannot be applied to individual terms.} Eliminating operators with field redefinitions will typically shuffle all kinds of contributions between many other operators. 

The dimension-six output at one-loop order is rather lengthy and is provided in the example notebook \texttt{Examples/E\_VLL\_model.nb} included in the public \matchete release along with the details of the comparison with Ref.~\cite{Carmona:2021xtq}. Instead of showing the full result, we illustrate the use of \mmaInlineCell[]{Input}{CovariantLoop} here. This \matchete routine provides the contribution from individual supertraces (or, equivalently, individual covariant loop topologies). For instance, to compute the contribution to the $(H^\dagger H)^3$ operator arising from the fermion hexagon graph (involving $3$ vector-like and $3$ SM lepton-doublet propagators), we simply run
\begin{mmaCell}{Input}
  CovariantLoop[\mmaDef{LUV}, \{Bar[\mmaDef{EE}], \mmaDef{l}, Bar[\mmaDef{EE}], \mmaDef{l}, Bar[\mmaDef{EE}], \mmaDef{l}\}] //NiceForm
\end{mmaCell}
\begin{mmaCell}{Output}
  \mmaFrac{1}{3} \!\!\(\hbar\)\!\! \mmaSup{\(\overline{yE}\)}{p}\!\! \mmaSup{\(\overline{yE}\)}{r}\!\! \mmaSup{\(\overline{yE}\)}{s}\!\! \mmaSup{yE}{p}\!\! \mmaSup{yE}{r}\!\! \mmaSup{yE}{s}\!\! \mmaFrac{1}{\mmaSup{ME}{2}}\!\! \mmaSub{\(\overline{H}\)}{i}\!\! \mmaSub{\(\overline{H}\)}{j}\!\! \mmaSub{\(\overline{H}\)}{j}\!\! \mmaSup{H}{i}\!\! \mmaSup{H}{j}\!\! \mmaSup{H}{k}\!\!
\end{mmaCell}
which coincides with the six NP Yukawa term in Eq.~(6.15) of Ref.~\cite{Carmona:2021xtq}. 

\subsection{Further examples and applications}\label{sec:otherNP}

Despite its limitations, this proof-of-concept version of \matchete already automates EFT matching computations for a wide class of weakly-coupled UV models. Models with an arbitrary particle content can be matched at tree-level, and automated one-loop matching for any model with heavy scalars and/or fermions (but no heavy vectors) is now possible. These matching steps can be performed up to arbitrary order in the heavy-mass expansion \textit{without requiring any further input from the user}. This is subject to constraints from the fast growth in computing time and memory requirements with increasing mass dimension, an area in which there is still room for improvement in future releases. While these functionalities certainly leave out many relevant models, they can already be used in multiple phenomenologically interesting applications. A non-exhaustive list of recent literature results for which \matchete could have been particularly useful include the matching of BSM models featuring:
\begin{itemize}
    \item[i)] Heavy scalars, like SM-singlets~\cite{Jiang:2018pbd,Haisch:2020ahr} (see Sec.~\ref{sec:singletscalar}), extra Higgses~\cite{Dawson:2022cmu}, electroweak triplets~\cite{Du:2022vso,Li:2022ipc}, flavorful triplets~\cite{de_Lima_2023}, or colorful new states~\cite{Gherardi:2020det,Gherardi:2020qhc,Dedes:2021abc}.
    \item[ii)] Heavy fermions, such as sterile neutrinos~\cite{Zhang:2021jdf,Ohlsson:2022hfl,Du:2022vso,de_Lima_2023}, other vector-like leptons~\cite{Carmona:2021xtq,Du:2022vso} (see Sec.~\ref{sec:VLLexample}), and vector-like quarks~\cite{Crivellin:2022fdf}.
    \item[iii)] Any combination of heavy scalars and fermions~\cite{Arnan:2016cpy,Crivellin:2021rbq,Liao:2022cwh,Guedes:2022cfy}.
\end{itemize}
As we discuss in the next section, we intend to expand these functionalities to cover an even wider range of applications.

Moreover, the automated simplification routines contained in \matchete, although not exhaustive, introduce entirely new capabilities to the EFT toolbox. While other tools~\cite{Criado:2019ugp,Gripaios:2018zrz,Fonseca:2017lem} deduce a suitable EFT basis, the simplification routines of \matchete automatically simplify an EFT with redundant operators and bring it to a form with mostly basis operators.\footnote{Some redundant operators are retained due to a lack of a more robust handling of flavor indices and an implementation of Fierz (and other four-dimensional) identities. As discussed in the next section, the latter is due to the appearance of evanescent operators, which need a dedicated treatment.} Despite some redundant operators persisting in some cases, the current implementation goes most of the way towards producing an EFT in a basis form. Failures to reach the basis with the current implementation are not critical, as on-/off-shell equivalence is preserved. To our knowledge, this is the first attempt at automatic simplification without any basis-specific hard-coded identities. These routines becomes particularly useful in studies where higher-dimensional operators are involved (see e.g. Refs.~\cite{Chala:2021wpj,de_Lima_2023,Banerjee:2022thk,Dawson:2022cmu,Chala:2023jyx} for recent literature examples), as the operator-reduction identities needed to obtain a basis are typically harder to implement due to the large proliferation of terms.

\section{Conclusions and Future Prospects}\label{sec:prospects}

    \begin{figure}
        \centering
        \includegraphics{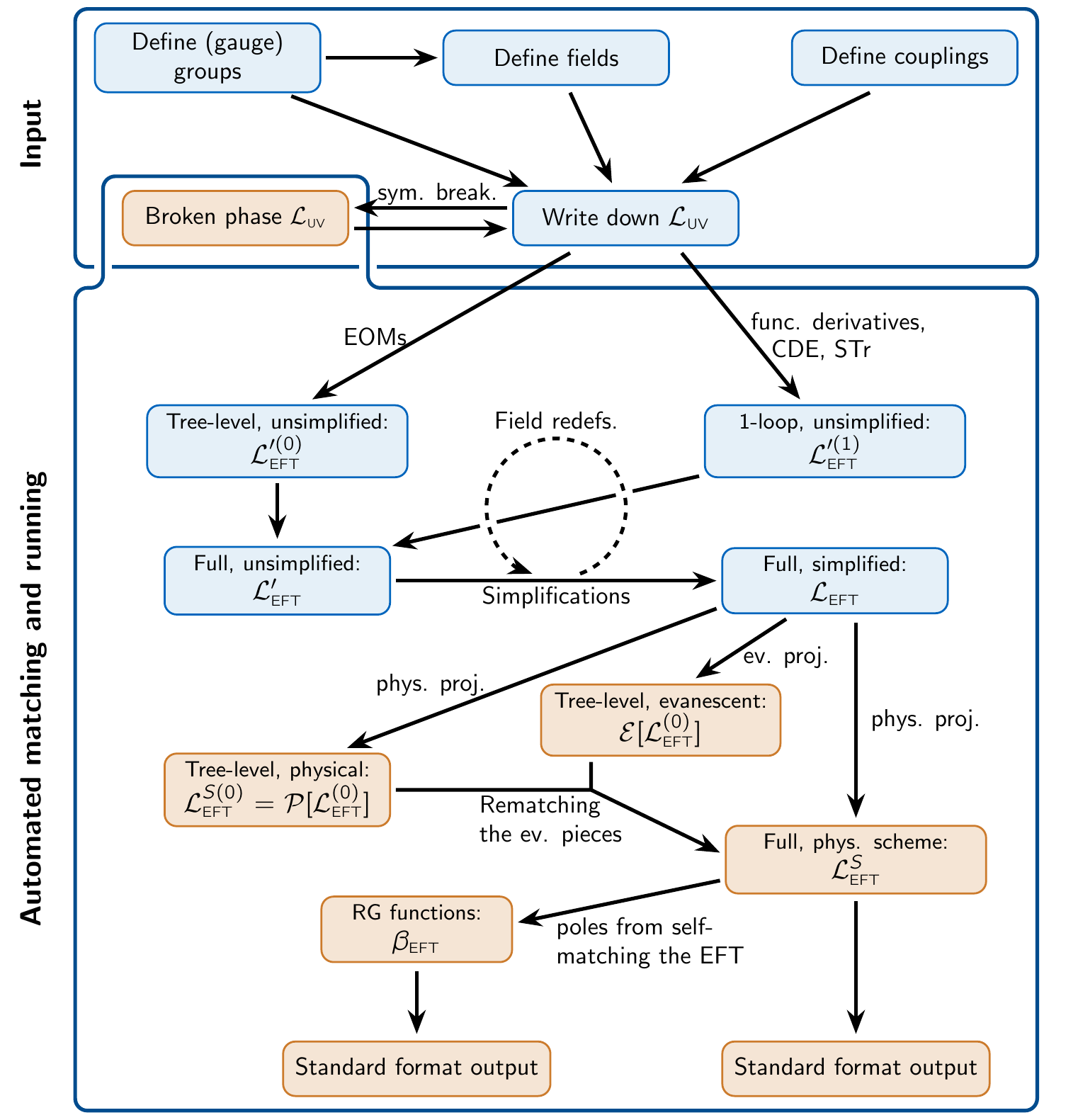}
        \caption{Roadmap for the future capabilities of \matchete. The workflow contained in the blue boxes are implemented in the proof-of-concept version, whereas the orange boxes are features expected in future releases. Here \textit{Standard format output} refers to both EFT basis identification and interfacing with other EFT tools.}
        \label{fig:matchete_extension_diagram} 
    \end{figure}

In this article, we have introduced the first version of \matchete and sketched out the workings of its routines. Already in this first version, it has great utility and versatility and can perform the matching of a wide range of UV models without any additional input for group theory or EFT bases: In its current form, \matchete is able to integrate out heavy scalars and fermions at the one-loop level, as well as heavy vectors at tree-level, with no restriction on the mass dimension of the effective theory, other than computing time limitations. As demonstrated in Sec.~\ref{sec:examples}, this already proves useful for a variety of concrete applications. Furthermore, the simplification routines can be used to automatically reduce Lagrangians to a close-to-basis form, even if the tool is not used for matching.

It is also clear that we can enhance the capabilities of \matchete even further. 
Our roadmap for future functionality includes addressing the following points:
\begin{itemize}
    \item Currently, the matching is performed in strictly $ d $ dimensions, which prevents EFT simplifications to the four-dimensional basis. We intend to implement the methods of Ref.~\cite{Fuentes-Martin:2022vvu} for defining a physical projection on the operator space and matching the remnant evanescent operator to the physical space.  
    \item After the implementation of routines for handling the evanescent operators, it will be possible to reduce EFT Lagrangians all the way to specific four-dimensional bases. The idea is to use this to obtain matching results as Wilson coefficients of known EFT bases, such as the Warsaw basis for the SMEFT or the LEFT basis of Ref.~\cite{Jenkins:2017jig}. Hence, it will be possible to interface the result with phenomenology packages through export in the \texttt{WCxf}~\cite{Aebischer:2017ugx} format. The interface with other phenomenology tools and/or commonly used formats, such as \texttt{UFO}~\cite{Degrande:2011ua}, would also be desirable.  
    \item The restriction of heavy states to scalars and fermions is the primary limitation of \matchete \texttt{v0.1.0}. In weakly-coupled theories, heavy vectors must arise from spontaneous symmetry breaking. This results in a complicated interplay between vectors, ghosts, and Goldstone bosons, especially in the background field gauge. So as to avoid having to derive and input all interactions manually, we wish to provide (semi-)automated methods to determine the broken phase Lagrangian.    
    \item With small changes to the matching procedure, it is possible to determine the EFT counterterms and, thereby, the RG functions. Implementing this functionality in \matchete will allow for finding the RG functions for intermediate-scale EFTs and vastly simplify sequential matching scenarios. 
\end{itemize}
In Figure~\ref{fig:matchete_extension_diagram} we show how the future functionalities fit into the \matchete workflow. The roadmap is our vision for the future of the \matchete package. It is of course subject to changes, as we determine what features are most important, or if the implementation of the functions becomes problematic.

Automated tools like \matchete have the potential to fundamentally change the workflow of BSM physics. They allow the practitioner to focus less on mechanical tasks and instead concentrate on finding answers to open questions in physics. Even with its current limitations, the proof of concept for \matchete already provides a valuable tool for NP phenomenology, and it demonstrates that functional methods offer a natural formulation of the matching (and RG running) procedure.

\subsection*{Acknowledgements}
We thank Jos\'e Santiago for his help with the cross-checks of the vector-like lepton example in Section~\ref{sec:VLLexample}. The authors are grateful to the Mainz Institute for Theoretical Physics (MITP) of the Cluster of Excellence PRISMA$^+$ (Project ID 39083149) for its hospitality and support. JF thanks Matthias Neubert and the Theoretical High Energy Physics Department at JGU Mainz for hospitality and support during his stay as a visitor. The work of JF has been supported by the Spanish Ministry of Science and Innovation (MCIN) and the European Union NextGenerationEU/PRTR under grant IJC2020-043549-I, by the MCIN grant PID2019-106087GB-C22, and by the Junta de Andaluc\'ia grants P21\_00199, P18-FR-4314 (FEDER) and FQM101. This research was supported in part by the Deutsche Forschungsgemeinschaft (DFG, German Research Foundation) through the Sino-German Collaborative Research Center TRR110 “Symmetries and the Emergence of Structure in QCD” (DFG Project-ID 196253076, NSFC Grant No. 2070131001, - TRR 110). JP and FW received funding from the European Research Council (ERC) under the European Union's Horizon 2020 research and innovation programme under grant agreement 833280 (FLAY), and by the Swiss National Science Foundation (SNF) under contract 200020-204428. The work of JP is supported by the U.S.\ Department of Energy (DOE) under award number~DE-SC0009919. The work of AET has received funding from the Swiss National Science Foundation (SNF) through the Eccellenza Professorial Fellowship ``Flavor Physics at the High Energy Frontier'' project number 186866. 

\appendix
\mmaSet{index=2}
\renewcommand{\thesection}{\Alph{section}}
\section{Defining the SM in Matchete}\label{app:SM}
To illustrate the ease of defining general quantum field theories in \matchete, we showcase how to input the SM Lagrangian. When using \matchete for a BSM theory, it is not necessary to repeat the SM input, as its definition can be loaded running the \mmaInlineCell[]{Input}{LoadModel["SM"]} routine, as shown in the examples in Sec.~\ref{sec:singletscalar} and~\ref{sec:VLLexample}.

In a first step, we use the \mmaInlineCell[]{Input}{DefineGaugeGroup} routine to define the SM gauge group $\mathrm{SU}(3)_c \times \mathrm{SU}(2)_L \times \mathrm{U}(1)_Y$ by
\begin{mmaCell}{Input}
   DefineGaugeGroup[SU3c, SU[3], gs, G,\newline    FundAlphabet -> \{"a","b","c","d","e","f"\},\newline    AdjAlphabet -> \{"A","B","C","D","E","F"\}]\newline DefineGaugeGroup[SU2L, SU[2], gL, W,\newline     FundAlphabet -> \{"i","j","k","l","m","n"\},\newline     AdjAlphabet -> \{"I","J","K","L","M","N"\}]\newline DefineGaugeGroup[U1Y, U1, gY, B]
\end{mmaCell}
where the groups are labeled \mmaInlineCell[]{Input}{SU3c}, \mmaInlineCell[]{Input}{SU2L}, and \mmaInlineCell[]{Input}{U1Y}, respectively. This automatically defines also the associated field-strength tensors, labeled \mmaInlineCell[]{Input}{\mmaDef{G}},~\mmaInlineCell[]{Input}{\mmaDef{W}}, and~\mmaInlineCell[]{Input}{\mmaDef{B}}, and the gauge couplings \mmaInlineCell[]{Input}{\mmaDef{gs}},~\mmaInlineCell[]{Input}{\mmaDef{gL}}, and~\mmaInlineCell[]{Input}{\mmaDef{gY}}. The optional arguments \mmaInlineCell[]{Input}{FundAlphabet} and \mmaInlineCell[]{Input}{AdjAlphabet} determine how fundamental and adjoint indices are displayed when using the \mmaInlineCell[]{Input}{NiceForm} routine.
Flavor indices and their printing style can be defined similarly:
\begin{mmaCell}{Input}
  DefineFlavorIndex[Flavor,3,IndexAlphabet->\{"p","r","s","t","u","v"\}]
\end{mmaCell}

Next, we define the field content of the SM using the \mmaInlineCell[]{Input}{DefineField} routine. We begin with the fermions:
\begin{mmaCell}{Input}
   DefineField[q, Fermion, Indices -> \{\mmaDef{SU3c}[fund],\mmaDef{SU2L}[fund],Flavor\},\newline     Charges -> \{U1Y[1/6]\}, Chiral -> LeftHanded, Mass -> 0]\newline DefineField[u, Fermion, Indices -> \{\mmaDef{SU3c}[fund], Flavor\},\newline    Charges -> \{U1Y[2/3]\}, Chiral -> RightHanded, Mass -> 0]\newline DefineField[d, Fermion, Indices -> \{\mmaDef{SU3c}[fund], Flavor\},\newline    Charges -> \{U1Y[-1/3]\}, Chiral -> RightHanded, Mass -> 0]\newline DefineField[\mmaUnd{l}, Fermion, Indices -> \{\mmaDef{SU2L}[fund], Flavor\},\newline    Charges -> \{U1Y[-1/2]\}, Chiral -> LeftHanded, Mass -> 0]\newline DefineField[e, Fermion, Indices -> \{Flavor\},\newline    Charges -> \{U1Y[-1]\}, Chiral -> RightHanded, Mass -> 0]
\end{mmaCell}
For example, the left-handed quark doublet is afterwards written in \matchete by typing \mmaInlineCell[]{Input}{\mmaDef{q}[a,i,p]}, where \mmaInlineCell[]{Input}{a}, \mmaInlineCell[]{Input}{i}, and~\mmaInlineCell[]{Input}{p} are the $\mathrm{SU}(3)_c$,  $\mathrm{SU}(2)_L$, and flavor indices, respectively, given in the order used in \mmaInlineCell[]{Input}{DefineField}.
The Higgs doublet is defined similarly:
\begin{mmaCell}{Input}
  DefineField[\mmaUnd{H}, Scalar, Indices -> \{\mmaDef{SU2L}[fund]\},\newline    Charges -> \{U1Y[1/2]\}, Mass -> 0];
\end{mmaCell}
The Higgs is defined as massless, because the mass parameter of the Higgs doublet is tachyonic. We will therefore include it manually as an interaction below.

The last missing definitions are for the (non-gauge) coupling constants for which we use the \mmaInlineCell[]{Input}{DefineCoupling} routine. The Yukawa matrices can be added with
\begin{mmaCell}{Input}
   DefineCoupling[Yu, Indices -> \{Flavor, Flavor\}]\newline DefineCoupling[Yd, Indices -> \{Flavor, Flavor\}]\newline DefineCoupling[Ye, Indices -> \{Flavor, Flavor\}]
\end{mmaCell}
Similarly, we can define the parameters of the Higgs potential by
\begin{mmaCell}{Input}
  DefineCoupling[\mmaUnd{\(\mu\)}, SelfConjugate -> True, EFTOrder -> 1]
  DefineCoupling[\mmaUnd{\(\lambda\)}, SelfConjugate -> True, EFTOrder -> 0]
\end{mmaCell}
where the option \mmaInlineCell[]{Input}{EFTOrder->1} specifies that the Higgs mass parameter \mmaInlineCell[]{Input}{\mmaUnd{\(\mu\)}} should be counted as having light-mass dimension one in the EFT power counting.

We can now write down the SM Lagrangian. Starting with the Yukawa interactions, we have
\begin{mmaCell}{Input}
  YukawaL = \mmaDef{Ye}[p,r] Bar[\mmaDef{l}[i,p]]**\mmaDef{e}[r] \mmaDef{H}[i]\newline    + \mmaDef{Yd}[p,r] Bar[\mmaDef{q}[a,i,p]]**\mmaDef{d}[a,r] \mmaDef{H}[i]\newline    + \mmaDef{Yu}[p,r] Bar[\mmaDef{q}[a,i,p]]**\mmaDef{u}[a,r] CG[eps[\mmaDef{SU2L}], \{i,j\}] Bar[\mmaDef{H}[j]];
\end{mmaCell}
The scalar potential is written as
\begin{mmaCell}{Input}
  HiggsPotential = -\mmaSup{\(\mu\)[]}{2}\,Bar[H[i]]H[i] + \mmaFrac{\(\lambda\)[]}{2}\,Bar[H[i]]H[i]Bar[H[j]]H[j];
\end{mmaCell}
Eventually, we can write the full SM Lagrangian
\begin{mmaCell}{Input}
   LSM = FreeLag[\mmaDef{q}, \mmaDef{u}, \mmaDef{d}, \mmaDef{l}, \mmaDef{e}, \mmaDef{H}, \mmaDef{G}, \mmaDef{W}, \mmaDef{B}]\newline    - PlusHc[\mmaDef{YukawaL}] - \mmaDef{HiggsPotential} //RelabelIndices;\newline LSM //HcSimplify //NiceForm
\end{mmaCell}
\begin{mmaCell}{Output}
  -\mmaFrac{1}{4}\mmaSup{\mmaSup{B}{\(\mu\nu\)}}{2} -\mmaFrac{1}{4}\mmaSup{\mmaSup{G}{\(\mu\nu\)A}}{2} -\mmaFrac{1}{4}\mmaSup{\mmaSup{W}{\(\mu\nu\)I}}{2} +\mmaSub{D}{\(\mu\)}\mmaSub{\(\overline{H}\)}{i}\mmaSup{D}{\(\mu\)}\mmaSup{H}{i} +\mmaSup{\(\mu\)}{2}\mmaSub{\(\overline{H}\)}{i}\mmaSup{H}{i} -\mmaFrac{1}{2}\(\lambda\)\mmaSub{\(\overline{H}\)}{i}\mmaSub{\(\overline{H}\)}{j}\mmaSup{H}{i}\mmaSup{H}{j} +i(\mmaSubSup{\(\overline{d}\)}{a}{p}\(\cdot\)\mmaSub{\(\gamma\)}{\(\mu\)}\mmaSub{P}{R}\(\cdot\)\mmaSub{D}{\(\mu\)}\mmaSup{d}{ap})\newline\newline  +i(\mmaSup{\(\overline{e}\)}{p}\(\cdot\)\mmaSub{\(\gamma\)}{\(\mu\)}\mmaSub{P}{R}\(\cdot\)\mmaSub{D}{\(\mu\)}\mmaSup{e}{p}) +i(\mmaSubSup{\(\overline{l}\)}{i}{p}\(\cdot\)\mmaSub{\(\gamma\)}{\(\mu\)}\mmaSub{P}{L}\(\cdot\)\mmaSub{D}{\(\mu\)}\mmaSup{l}{ip}) +i(\mmaSubSup{\(\overline{q}\)}{ai}{p}\(\cdot\)\mmaSub{\(\gamma\)}{\(\mu\)}\mmaSub{P}{L}\(\cdot\)\mmaSub{D}{\(\mu\)}\mmaSup{q}{aip}) +i(\mmaSubSup{\(\overline{u}\)}{a}{p}\(\cdot\)\mmaSub{\(\gamma\)}{\(\mu\)}\mmaSub{P}{R}\(\cdot\)\mmaSub{D}{\(\mu\)}\mmaSup{u}{ap})\newline\newline  +\big(-\mmaSup{Ye}{rp}\mmaSup{H}{i}(\mmaSubSup{\(\overline{l}\)}{i}{r}\(\cdot\)\mmaSub{P}{R}\(\cdot\)\mmaSup{e}{p}) -\mmaSup{Yd}{rp}\mmaSup{H}{i}(\mmaSubSup{\(\overline{q}\)}{ai}{r}\(\cdot\)\mmaSub{P}{R}\(\cdot\)\mmaSup{d}{ap}) -\mmaSup{Yu}{rp}\mmaSub{\(\overline{H}\)}{i}(\mmaSubSup{\(\overline{q}\)}{aj}{r}\(\cdot\)\mmaSub{P}{R}\(\cdot\)\mmaSup{u}{ap})\mmaSup{\(\varepsilon\)}{ji} + H.c.\big)
\end{mmaCell}
where we used \mmaInlineCell[]{Input}{RelabelIndices} to canonically relabel all indices in the Lagrangian, \mmaInlineCell[]{Input}{PlusHc} to include the Hermitian conjugate of the Yukawa Lagrangian defined above, and \mmaInlineCell[]{Input}{HcSimplify} to collect hermitian conjugated terms for the printing. 

We can also use the \mmaInlineCell[]{Input}{CheckLagrangian} routine to perform a series of checks on the Lagrangian, ensuring for example that \mmaInlineCell[]{Input}{\mmaDef{LSM}} is Hermitian, gauge-invariant, is free of gauge anomalies, and more:
\begin{mmaCell}{Input}
  CheckLagrangian[\mmaDef{LSM}]
\end{mmaCell}
\begin{mmaCell}{Output}
  True
\end{mmaCell}
These checks are also internally applied by the \mmaInlineCell[]{Input}{Match} routine, before performing the matching, which can only be done for Lagrangians passing all tests of \mmaInlineCell[]{Input}{CheckLagrangian}.

The addition of further BSM fields, and couplings works analogously. For BSM theories with extended gauge symmetries, the BSM Lagrangian would have to be provided to \matchete in the broken phase, where only the residual SM gauge symmetry remains unbroken. The complexity of the resulting Lagrangian with ghosts, heavy vectors, and gauge fixing, and the precise relations among them needed to obtain meaningful results is why this is not presently supported.

\section*{References}
{
\bibliography{References}
}

\end{document}